# $^{60}$Fe-$^{60}$Ni chronology of core formation in Mars

Haolan Tang,[1,*,†] Nicolas Dauphas[1]

[1]Origins Lab, Department of the Geophysical Sciences and Enrico Fermi Institute, The University of Chicago, 5734 South Ellis Avenue, Chicago IL 60637;

[*]To whom correspondence should be addressed. E-mail: haolantang@ucla.edu

[†] Present address: Ion Probe Group, Department of Earth and Space Sciences, University of California, Los Angeles, 595 Charles E.Young Drive East, Los Angeles CA, 90095

Abstract

The timescales of accretion, core formation, and magmatic differentiation in planetary bodies can be constrained using extinct radionuclide systems. Experiments have shown that Ni becomes more siderophile with decreasing pressure, which is reflected in the progressively higher Fe/Ni ratios in the mantles of Earth, Mars and Vesta. Mars formed rapidly and its mantle has a high Fe/Ni ratio, so the $^{60}$Fe-$^{60}$Ni decay system ($t_{1/2}$=2.62 Myr) is well suited to establish the timescale of core formation in this object. We report new measurements of $^{60}$Ni/$^{58}$Ni ratios in bulk SNC/martian (Shergotty-Nakhla-Chassigny) meteorites and chondrites. The difference in $\varepsilon^{60}$Ni values between SNC meteorites and the building blocks of Mars assumed to be chondritic (55 % ordinary chondrites +45% enstatite chondrites) is +0.028±0.023 (95% confidence interval). Using a model of growth of planetary embryo, this translates into a time for Mars to have reached ~44 % of its present size of $1.9^{+1.7}_{-0.8}$ Myr with a strict lower limit of 1.2 Myr after solar system formation, which agrees with a previous estimate based on $^{182}$Hf-$^{182}$W systematics. The presence of Mars when planetesimals were still being formed may have influenced the formation of chondrules through bow shocks or by inducing collisions between dynamically excited planetesimals. Constraints on the growth of large planetary bodies are scarce and this is a major development in our understanding of the chronology of Mars.

## 1. Introduction:

The timescales of accretion, core formation, and mantle differentiation in planetary bodies are key observables that can help us test theories regarding the origin and early evolution of asteroids and planets (see recent reviews by Wadhwa et al., 2006; Dauphas and Chaussidon, 2011). The extinct $^{60}$Fe-$^{60}$Ni system ($t_{1/2}$ = 2.62 Myr; Rugel et al., 2009) is a potentially powerful tool to constrain the timescale of core formation in these bodies because (1) Ni is more siderophile than Fe, therefore the Fe/Ni ratio left in the mantle after core segregation can be strongly fractionated relative to chondrites and (2) core formation in planetesimals and embryos is thought to have occurred within the first several million years of the formation of the solar system, when $^{60}$Fe was still extant. Much work has been done previously to estimate the timescales of core formation in various bodies using $^{182}$Hf-$^{182}$W systematics ($t_{1/2}$=9 Myr) but ambiguities remain on several aspects of early solar system chronology (see Kleine et al., 2009 for a review). For example, W isotope measurements of SNC (Shergotty-Nakhla-Chassigny) meteorites constrain the time after formation of the solar system for the core of Mars to have reached approximately half of its present size to approximately $1.8^{+0.9}_{-1.0}$ Myr (Dauphas and Pourmand, 2011; Kobayashi and Dauphas, 2013). However, this estimate is really an upper-limit (Mars could have accreted more rapidly than that), as SNC meteorites contain variable excess $^{182}$W produced by early silicate differentiation and the accretion timescale was calculated using the least radiogenic $^{182}$W values (Kleine et al., 2004a, 2009; Foley et al., 2005). The $^{60}$Fe-$^{60}$Ni system has seldom been used to constrain early solar system chronology and most studies have focused instead on establishing the initial abundance of $^{60}$Fe in meteorites, with important implications for the astrophysical context of solar system formation. Indeed, high $^{60}$Fe/$^{56}$Fe ratios obtained by measuring nickel isotopic compositions as well as Fe/Ni ratios by secondary ion mass spectrometers (SIMS) in chondrites were initially taken as fingerprints of the injection of fresh nucleosynthetic products into the solar system by the explosion of a nearby supernova (Tachibana et al., 2003, 2006; Mostefaoui et al., 2004, 2005; Guan et al., 2007; Marhas and Mishra, 2012; Mishra et al., 2010; Mishra and Chaussidon, 2012). Other studies of achondrites and chondritic constituents found lower $^{60}$Fe/$^{56}$Fe ratios that were interpreted to reflect late disturbance (Chen et al., 2013) or heterogeneous distribution of $^{60}$Fe in the early solar system (Sugiura et al., 2006; Quitté et al., 2010, 2011). Recently, Tang and Dauphas (2012) determined by MC-ICPMS the Ni isotopic compositions of bulk angrites, bulk HEDs, bulk chondrites, mineral separates of quenched angrite D'Orbigny, Gujba chondrules, and chondrules and mineral separates from unequilibrated ordinary chondrites (Semarkona and NWA 5717). In all objects analyzed (chondrites and achondrites). They found uniformly low $^{60}$Fe/$^{56}$Fe ratios corresponding to an initial ratio at the time of condensation of the first solids in the solar nebula (calcium-aluminum-rich inclusions, CAIs) of $(1.15\pm0.26)\times10^{-8}$. Strengthening the case for a uniform $^{60}$Fe/$^{56}$Fe ratio, they also reported high-precision $^{58}$Fe isotope measurements and detected no anomaly in this isotope. The rationale for measuring this isotope of iron is that $^{58}$Fe and $^{60}$Fe are produced together by neutron captures in supernovae, so any heterogeneity in $^{60}$Fe should be accompanied by heterogeneity in $^{58}$Fe (Dauphas et al., 2008). The contrapositive of this statement is that $^{58}$Fe homogeneity implies $^{60}$Fe homogeneity at a level that rules out the highest $^{60}$Fe/$^{56}$Fe ratios measured by SIMS in chondrules. Spivack-

Birndorf et al. (2012) also concluded that $^{60}$Fe was present in low abundance in the early solar system.

The study of Tang and Dauphas (2012) puts $^{60}$Fe-$^{60}$Ni on solid footings to investigate early solar system chronology. Tang and Dauphas (2012) were thus able to constrain the time of core formation on Vesta to $3.7^{+2.5}_{-1.7}$ Myr after solar system formation, using Ni isotope measurements of HED (Howardite-Eucrite-Diogenite) meteorites. Here, we report Ni isotope measurements of SNC meteorites and chondrites that provide new constraints on the timescale of core formation in Mars.

The dynamical context of Mars' accretion is the subject of much work and speculation. Terrestrial planets are thought to have formed through collisions between Moon to Mars-size planetary embryos over a duration of several tens of million years (Chambers and Wetherill, 1998). Modeling of terrestrial planet formation can reproduce the size and accretion timescale of Earth but the same simulations fail to explain the small size of Mars (Wetherill, 1991; Raymond et al., 2009). One possibility is that Mars grew rapidly into a planetary embryo by accretion of planetesimals and evaded collisions with other embryos during the subsequent stage of chaotic growth, as was experienced by Earth (Chambers and Wetherill, 1998; Chambers, 2004; Kobayashi and Dauphas, 2013). The short accretion timescale of Mars is indeed consistent with this view (Dauphas and Pourmand, 2011). The rapid growth and small mass of Mars can be achieved in the framework of the grand tack scenario, whereby Jupiter migrated inward by type II migration, truncated the inner disk at 1 AU, at which point it entered in a 2:3 resonance with Saturn and started migrating outward (Walsh et al., 2011; Pierens and Raymond, 2011). Accordingly, Mars would have been scattered early on to its present location and would have stopped growing significantly afterwards. Mars may thus be the sole representative of a generation of planetary bodies that preceded the formation of other terrestrial planets. However, the question of how fast Mars formed remains opened because $^{182}$Hf-$^{182}$W only constrains its accretion to have occurred in the first few million years of solar system formation, a period of time when the solar protoplanetary disk changed rapidly. For example, it was suggested that chondrules could have formed by bow shocks from planetary embryos like Mars (Morris et al., 2012; Boley et al., 2013) but the relative chronology of embryo *vs.* chondrule formation remains uncertain. Refining the chronology of Mars formation is also important for constraining models of embryo growth. Kobayashi and Dauphas (2013) used the mass and accretion time of Mars as constraints in statistical simulations of embryo growth to conclude that Mars most likely formed in a massive disk from relatively small planetesimals, an idea that can be tested further by reducing uncertainties in the growth history of Mars.

In order to establish the timescale of Mars formation using the $^{60}$Fe-$^{60}$Ni extinct chronometer, we have measured the Ni isotopic compositions of 5 martian (SNC) meteorites and 11 chondrites that are thought to represent the building blocks of Mars. The Ni isotopic results give a timescale of $1.9^{+1.7}_{-0.8}$ Myr for core formation and accretion on Mars, with a robust lower limit of 1.2 Myr.

**2. Methodology**

**2.1 Sample preparation, digestion, and chemical separation**

All the chemistry was performed under clean laboratory conditions at the Origins Lab of the University of Chicago. Optima grade HF, reagent grade acetone, and double distilled HCl and HNO$_3$ were used for digestion and column chromatography. Millipore Milli-Q water was used for acid dilution. Bulk chondrites were selected to help estimate the Ni isotopic composition of bulk Mars and comprise 5 carbonaceous chondrites (Allende, Murchison, Mighei, Vigarano, Orgueil, Leoville) and 6 H-chondrites (Bath, Bielokrynitschie, Kesen, Kernouvé, Ochansk, Ste. Marguerite). These samples were provided by the Field Museum. The martian meteorites (Shergotty, Chassigny, Nakhla, Zagami, Lafayette) were provided by the Smithsonian Institution. All samples were crushed into powder in an agate mortar before digestion. To assess data quality and make sure that no analytical artifacts were present, terrestrial standards were processed and analyzed together with the meteorite samples.

Sample digestion and chemical separation are described by Tang and Dauphas (2012). Chondrites and martian meteorites weighing 8 to 160 mg were digested in 5-30 ml HF-HNO$_3$ (in a 2:1 volume ratio) in Teflon beakers placed on a hot plate at ~90 ˚C for 5-10 days. The solutions were subsequently evaporated to dryness and re-dissolved in a 5-30 mL mixture of concentrated HCl-HNO$_3$ with a volume ratio of 2:1. For the samples that were not digested completely, the residues were separated by centrifugation and digested in HF-HNO$_3$ (2:1 volume ratio) and HNO$_3$ using Parr bombs at 90 °C for 5-10 days until complete digestion. The solutions were dried down and the residues taken back in solution with a minimum amount of concentrated HCl (~11 M) for loading on the first column. In order to obtain sufficiently clean Ni cuts for isotopic measurements, chemical separation of Ni from matrix elements and isobars was done using three steps of liquid chromatography.

U/TEVA cartridge (Horwitz et al., 1992) was used for the first chemistry step to get rid of Ti, Co, Zr and Fe. The column (2 mL volume, 2.5 cm length, 1 cm diameter) was pre-cleaned with 10 mL water, 15 mL 0.4 M HCl, 15 mL 4 M HCl and was then conditioned with 10 mL of concentrated HCl. The sample solution was loaded onto the column in 5-10 mL 10 M HCl. The load solution was collected in clean Teflon beakers and an additional 10 mL of concentrated HCl was passed through the resin and collected in the same beaker. This eluate contained Ni together with Na, Mg, Ca and other matrix elements. After drying down, the Ni elution cut from the first column chemistry was re-dissolved in 5 mL of a mixture of 20 % 10 M HCl-80 % acetone (by volume) and loaded onto 5 mL (40 cm length, 0.4 cm diameter) pre-cleaned Bio-Rad AG50-X12 200-400 mesh hydrogen-form resin in a Teflon column, previously conditioned with 10 mL 20 % 10 M HCl-80 % acetone. After loading the sample solution and rinsing with 30 mL 20 % 10 M HCl – 80% acetone mixture to eliminate Cr and any remaining Fe, Ni was collected by eluting 150 mL of the HCl-acetone mixture into a jar containing 30 mL $H_2O$ to dilute HCl and stabilize Ni in the eluate. In those conditions, Mg, Na, Ca, and other matrix elements were retained on the resin (Strelow et al., 1971; Tang and Dauphas, 2012). The collected Ni solution was evaporated at moderate temperature (<90 °C) under a flow of $N_2$ to avoid the formation of organic complexes with acetone and accelerate evaporation. After evaporation, the Ni fraction was dissolved in 1 mL of aqua regia (1:3 $HNO_3$:HCl) to remove any organic residue formed during evaporation. This HCl-acetone column was repeated five times to ensure thorough separation of major rock forming element Mg from Ni, two elements that are notoriously difficult to separate. Zinc is a significant interference on low abundance isotope $^{64}$Ni. It was removed using a third column filled with 1 mL (2 cm length, 0.8 cm diameter) Bio-Rad AG1W-X8 anionic ion exchange resin in 8 M HBr medium (Moynier et al., 2006). Nickel was eluted in 8 mL 8 M HBr, whereas Zn was retained on the resin.

The entire procedural blanks were ~35 ng for Ni isotope composition measurements, negligible compared to the amounts of Ni in the samples. Nickel yield of the entire procedure was 90-100 %.

### 2.2 Mass spectrometry

All measurements were performed at the Origins Laboratory of the University of Chicago using a Neptune MC-ICPMS equipped with an OnTool Booster 150 (Pfeiffer) interface jet pump. Jet sampler and X skimmer cones were used. The samples were re-dissolved in 0.3 $HNO_3$ and introduced into the mass spectrometer with Ar + $N_2$ using an Aridus II desolvating nebulizer at an uptake rate of ~ 100 μL/min. The instrument sensitivity for $^{58}Ni$ was 100 V/ppm. One analysis consisted of 25 cycles, each acquisition lasting for 8.4 s. During a session, each sample solution was measured 13 times bracketed by SRM 986. A small isobaric interference from the least abundant isotope of iron, $^{58}Fe$, on the most abundant isotope of nickel, $^{58}Ni$, was corrected by monitoring $^{57}Fe$ (the $^{58}Fe/^{58}Ni$ ratio was always less than 0.00005 in our measurements). All isotopes were measured with Faraday cups with $10^{11}$ Ω resistance amplifiers. Background was subtracted using an on-peak zero procedure. Internal normalization was used to correct mass-dependent isotopic fractionation by fixing $^{61}Ni/^{58}Ni$ to 0.016720 or $^{62}Ni/^{58}Ni$ to 0.053389 (Gramlich et al., 1989) using the exponential law (Maréchal et al., 1999).

In the following, we only discuss results based on the normalization to $^{61}$Ni/$^{58}$Ni because: (1) it reduces the spread in $\varepsilon^{60}$Ni values of chondrites (Table 1) arising from the presence of nucleosynthetic anomalies on Ni isotopes and (2) although Ti in the samples have been eliminated completely, there is still a potential interference on $^{62}$Ni coming from titanium oxide ($^{46}$Ti$^{16}$O) in the acid background (Tang and Dauphas, 2012).

Approximately 20-25 % of the original sample solutions were kept as safety aliquots and for Fe/Ni ratio measurements by MC-ICPMS using both the bracketing and standard addition techniques (see Tang and Dauphas 2012 for details). The standard addition method largely eliminates possible matrix effects in elemental analyses (*e.g.,* Harris, 2011). The Fe/Ni ratios measured by standard addition were compared with Fe/Ni ratios measured by simple sample-standard bracketing and the values were identical within uncertainties. Fe/Ni ratios in terrestrial standards were all within 3 % of their reference values, demonstrating the accuracy of our measurements.

The uncertainties on Fe and Ni concentrations are those given by Isoplot for the *x*-intercept of the standard addition data. Fe/Ni ratios in Table 1 were obtained by the standard addition technique.

### 3. Results

Table 1 shows the Ni isotopic compositions and Fe/Ni ratios measured in meteorites and terrestrial rock standards. Below, we focus our discussion on results reported using $^{61}$Ni/$^{58}$Ni internal normalization. Terrestrial standards passed through the same column chemistry as meteoritic samples have normal Ni isotopic ratios, attesting to the accuracy of the measurements. Note that $^{64}$Ni was not measured in several samples with very low Ni contents (BE-N, BHVO-2, DNC-1, and all SNC meteorites).

The Ni isotope measurements of bulk chondrites agree well with previous studies reporting the presence of small isotopic anomalies of nucleosynthetic origin for Ni isotopes in meteorites (Dauphas et al., 2008; Regelous et al., 2008; Steele et al., 2012). $\varepsilon^{62}$Ni and $\varepsilon^{64}$Ni isotopic anomalies in chondrites are correlated with each other (Fig. 1). Steele et al. (2012) measured the Ni isotopic composition of meteoritic materials using a double spike technique and concluded that the cause of $^{62}$Ni and $^{64}$Ni isotopic variations was most likely variations in the neutron-poor isotope $^{58}$Ni, which the authors ascribed to incomplete mixing of products of supernova nucleosynthesis in the solar protoplanetary disk. $\varepsilon^{62}$Ni and $\varepsilon^{64}$Ni are also correlated with $\varepsilon^{54}$Cr anomalies (Fig. 2A; $\varepsilon^{54}$Cr=7.430×$\varepsilon^{62}$Ni+0.055), the carrier of which has been identified as $^{54}$Cr-rich nanooxide/nanospinels of supernova origin (Dauphas et al, 2010; Qin et al., 2011). More work is needed to identify the presolar carrier of Ni isotopic anomalies in meteorites. $\varepsilon^{60}$Ni values in chondrites also show departure from terrestrial composition. These variations in the $^{60}$Ni isotopic composition of bulk chondrites are unlikely to be related to $^{60}$Fe because all bulk chondrites have similar Fe/Ni ratios yet distinct $\varepsilon^{60}$Ni values, which is impossible to explain if the $^{60}$Fe/$^{56}$Fe ratio was uniform (Tang and Dauphas, 2012). Furthermore, samples with low $\varepsilon^{60}$Ni have normal $\varepsilon^{58}$Fe, indicating a decoupling between variations in $^{60}$Ni isotopic anomalies and neutron-rich isotopes $^{58}$Fe and $^{60}$Fe. Isotopic anomalies in $^{60}$Ni do not correlate with $^{62}$Ni or $^{64}$Ni, calling for the presence of a third nucleosynthetic component in the solar system Ni isotope mix.

Within the general groups of ordinary and enstatite chondrites, the different samples have uniform Ni isotopic compositions. Combining the results presented in Table 1 and Figure 3 with previously published data (Dauphas et al., 2008; Regelous et al., 2008;

Steele et al., 2012), we estimate the weighted average $\varepsilon^{60}Ni$ isotopic values of ordinary and enstatite chondrites to be -0.048±0.008 (n=13) and -0.019±0.010 (n=5), respectively (see Table S1 in Appendix B for a compilation). All uncertainties are 95 % confidence intervals of the averages and were calculated using the Isoplot software.

Regardless of the variability in $^{142}Nd$ and $^{182}W$ anomalies among SNC meteorites (Kleine et al., 2004a, 2009; Foley et al., 2005), the Ni isotopic composition of five SNC meteorites (Shergotty, Chassigny, Nakhla, Zagami, Lafayette) is constant and identical to the terrestrial value (Fig. 3). No correlation is found between the Ni isotopic composition and Fe/Ni ratios of the different SNC meteorites (Fig. 4). The weighted average $\varepsilon^{60}Ni$ value of SNC meteorites is -0.010 ±0.022 (MSWD=0.034).

## 4. Discussion

Angrites and HED meteorites show variations in $\varepsilon^{60}Ni$ linearly correlated with Fe/Ni ratio, indicating global silicate differentiation while $^{60}Fe$ was still alive. In contrast, the uniform $\varepsilon^{60}Ni$ value found in all SNC meteorites, indicates that the magmatic differentiation events that fractionated Fe/Ni ratios in SNC meteorites must have taken place after $^{60}Fe$ had fully decayed. The inferred initial ratio of $(-0.01\pm1.19)\times10^{-9}$ at the time when Fe-Ni system was closed translates into a minimum age for Fe/Ni fractionation of 8.6 Myr after solar system formation, assuming $(^{60}Fe/^{56}Fe)_0=(1.15\pm0.26)\times10^{-8}$ as the initial ratio at the time of CAI formation (Fig. 4; Tang and Dauphas, 2012). This timescale is fully consistent with the ~40-100 Myr formation interval for the shergottite mantle sources, as derived from coupled $^{142,143}Nd$ systematics (Foley et al., 2005, Caro et al., 2008, Debaille et al., 2007), and with the 2-stage $^{146}Sm$-$^{142}Nd$ model age for the nakhlites-Chassigny group of ~25 Myr (Harper et al.,

1995, Foley et al., 2005). Some of the Fe/Ni fractionation recorded in SNC meteorites must reflect mineral/melt fractionation during partial melting of the martian mantle and subsequent differentiation of the magmas, which took place at least 0.5 Gyr after solar system formation (Lapen et al., 2010), well after complete decay of $^{60}$Fe. On the other hand, both the terrestrial and martian mantles are known to host heterogeneities in decay products of short-lived nuclides $^{146}$Sm and $^{182}$Hf (Harper et al., 1995; Lee and Halliday, 1997; Jagoutz et al., 2003; Kleine et al., 2004a; Foley et al., 2005; Caro et al., 2008; Debaille et al., 2007; Willbold et al., 2011; Touboul et al., 2012). No such heterogeneity was found for $^{60}$Fe-$^{60}$Ni in the martian or terrestrial mantles. The average $\varepsilon^{60}$Ni of all SNC meteorites measured in this study is taken as representative of the $\varepsilon^{60}$Ni value of the bulk Martian mantle ($\varepsilon^{60}$Ni =-0.010 ±0.022).

The time of core formation on Mars has previously been estimated using the $^{182}$Hf-$^{182}$W chronometer. A difficulty however is that different martian meteorites have variable $\varepsilon^{182}$W and $\varepsilon^{142}$Nd values, reflecting early Hf/W and Sm/Nd fractionation presumably associated with magma ocean crystallization (Kleine et al., 2004a, 2009; Foley et al., 2005). The $\varepsilon^{182}$W values in SNC meteorites range from +0.3 to +3 (Foley et al., 2005; Kleine et al., 2004a, 2009), the highest values being characteristic of Nakhlites that are more enriched and oxidized than other SNC meteorites and may bear the signature of a crustal component. Conservatively, one can use the lowest $\varepsilon^{182}$W of +0.4 ±0.2 measured in Shergottites and the Hf/W ratio of the martian mantle (3.51 ± 0.45; Dauphas and Pourmand, 2011) to estimate the age for core formation on Mars. Using this approach, Dauphas and Pourmand (2011) estimated that Mars accreted approximately 44 % of its present mass in $1.8^{+0.9}_{-1.0}$ Myr. As discussed in the introduction, this value is really an upper-limit, as the $^{182}$W isotopic composition of the martian mantle could be more radiogenic than $\varepsilon^{182}$W=+0.4 ±0.2.

The Hf-W approach takes advantage of the fact that core formation led to a drastic fractionation of Hf and W, as Hf is strongly lithophile and W is moderately siderophile. Similarly, core formation in the terrestrial planets fractionated Fe from Ni in the silicate mantle. If Mars' core formed early, while $^{60}$Fe was still alive, the high Fe/Ni ratio of the martian mantle should have produced excess radiogenic $^{60}$Ni relative to bulk Mars, taken to be chondritic. The ingredients to calculate a timescale of core formation on Mars are the Fe/Ni ratios and $\varepsilon^{60}$Ni values of the mantle of Mars and chondrites.

The $\varepsilon^{60}$Ni value of the martian mantle is that obtained here from measurements of SNC meteorites, *i.e.*, $\varepsilon^{60}$Ni$_{mantle}$= -0.010 ±0.022 (the error bar is the 95 % confidence interval of the weighted mean). The Fe/Ni ratio of the bulk mantle of Mars cannot be measured directly in SNC meteorites because Fe and Ni could be fractionated during melting and crystallization after $^{60}$Fe was extinct. Instead, the Fe/Ni ratio of the bulk mantle of Mars was estimated by Warren (1999) based on MgO-NiO and MgO-FeO correlations to be ~350, which corresponds to a $^{56}$Fe/$^{58}$Ni ratio of ~472 (also see Wänke and Dreibus, 1988).

As discussed in Sect. 3, chondrites show variations in Ni isotopic ratios (not only $\varepsilon^{60}$Ni but also $\varepsilon^{62}$Ni and $\varepsilon^{64}$Ni) that cannot be ascribed to radiogenic ingrowth but must reflect instead the presence of planetary scale nucleosynthetic anomalies, as has been documented previously for other elements such as Mo (Dauphas et al., 2002a, b; Burkhardt et al., 2011). In order to estimate the $\varepsilon^{60}$Ni value of bulk Mars, one must know the nature of the building blocks that made Mars. The average $\Delta^{17}$O, $\varepsilon^{50}$Ti, $\varepsilon^{54}$Cr, $\varepsilon^{62}$Ni, and $\varepsilon^{92}$Mo values of carbonaceous (CC), ordinary (OC), and enstatite (EC) chondrites are compiled in Figure 2 (Cr: Qin et al., 2010; Trinquier et al., 2007; Ti: Trinquier et al. 2009; Zhang et al., 2012; Mo: Dauphas et al., 2002a, b; Burkhardt et al., 2011; Ni: Dauphas et al., 2008; Steele et al., 2012; Tang and Dauphas, 2012; O: Clayton et al., 1983, 1984, 1991, 1999; Weisberg et al., 2001; Tables S1-S5). Warren (2011) showed that in any case, carbonaceous chondrites contributed less than 18 %. The mixture advocated by Sanloup et al. (1999) (55 % H + 45 % EH) can reproduce the $\Delta^{17}$O, $\varepsilon^{50}$Ti, $\varepsilon^{54}$Cr, $\varepsilon^{62}$Ni, and $\varepsilon^{92}$Mo values measured in SNC meteorites while that of Lodders and Fegley (1997) does not (85 % H + 11 % CV + 4 % CI; Fig. 2). Therefore, the bulk $\varepsilon^{60}$Ni isotopic composition of Mars is calculated using the following equation:

$$\varepsilon^{60}\text{Ni}_{\text{Bulk Mars}} = (\varepsilon^{60}\text{Ni}_{\text{OC}} X_H [\text{Ni}]_H + \varepsilon^{60}\text{Ni}_{\text{EC}} X_{EH} [\text{Ni}]_{EH})/(X_H [\text{Ni}]_H + X_{EH} [\text{Ni}]_{EH}) \quad (2),$$

where X is the weight fraction of the various chondrite types in bulk Mars (55% H+45% EH; Sanloup et al., 1999) and [Ni] is the Ni concentration (Wasson and Kallemeyn, 1988). For $\varepsilon^{60}$Ni, we take the average values for OC and EC as proxies for H and EH because meteorites within these groups have uniform Ni isotopic compositions. The resulting Ni isotopic composition of bulk Mars is $\varepsilon^{60}$Ni = -0.038±0.006. Using the mixing proportions proposed by Lodders and Fegley (1997), the $\varepsilon^{60}$Ni value of bulk Mars would have been -0.056±0.008, which is slightly lighter than the value adopted here. The Fe/Ni ratio of bulk Mars is calculated using the compilation of Wasson and Kallemeyn (1988) and the following mass-balance equation (55 % H+45 % EH):

$$\left(\frac{\text{Fe}}{\text{Ni}}\right)_{\text{Bulk Mars}} = \left[\left(\frac{\text{Fe}}{\text{Ni}}\right)_H X_H [\text{Ni}]_H + \left(\frac{\text{Fe}}{\text{Ni}}\right)_{EH} X_{EH} [\text{Ni}]_{EH}\right]/(X_H [\text{Ni}]_H + X_{EH} [\text{Ni}]_{EH}). \quad (3)$$

The Fe/Ni ratio of bulk Mars is therefore 17.0. The Fe/Ni ratio of the Martian core is not known directly but can be obtained by mass-balance,

$$\left(\frac{Fe}{Ni}\right)_{\text{Mars core}} = \frac{(Fe/Ni)_{\text{Bulk Mars}}[Ni]_{\text{Bulk Mars}}M_{\text{Bulk Mars}} - (Fe/Ni)_{\text{Mars mantle}}[Ni]_{\text{Mars mantle}}M_{\text{Mars mantle}}}{[Ni]_{\text{Bulk Mars}}M_{\text{Bulk Mars}} - [Ni]_{\text{Mars mantle}}M_{\text{Mars mantle}}}. \quad (4)$$

This yields a Fe/Ni ratio for Mars' core of 9.79.

To translate Ni isotope measurements of SNC meteorites into age constraints, a proper model of core formation must be used. A reasonable assumption to make is that core formation tracks planetary accretion, meaning that whenever mass is added to Mars, metal is removed into the core in proportion to the present core/mantle ratio. To parameterize the growth of Mars, Dauphas and Pourmand (2011) used some formalism relevant to oligarchic growth of embryos from planetesimals (Thommes et al., 2001; Chambers, 2006),

$$\frac{M_{\text{Mars}}(t)}{M_{\text{Mars}}} = \tanh^3(t/\tau), \quad (5)$$

where t is counted from the formation of the solar system, and $\tau$ is the accretion timescale. At $t = \tau$, the size of embryo is $\tanh^3(1) = 44\%$ of its present size. More recently, Kobayashi and Dauphas (2013) and Morishima et al. (2013) used more realistic model simulations (statistical or N-body) to investigate the influence of the accretion history of Mars on $^{182}$Hf-$^{182}$W model age. Much can be learned from these sophisticated simulations but the formalism used by Dauphas and Pourmand (2011) captures the most salient features of embryo growth in a single parameter ($\tau$), so this formalism is used here.

Another point that deserves consideration is the question of knowing whether metal in the incoming bodies was able to equilibrate fully with the whole mantle of protoMars (Halliday, 2004; Nimmo and Agnor, 2006; Kleine et al., 2004b; 2009; Nimmo et al., 2010; Rudge et al., 2010; Dahl and Stevenson, 2010; Deguen et al., 2011). At one extreme, one could imagine a model whereby no equilibration occurs and the metal of the incoming bodies sinks directly to the core, without exchange with the mantle. In this situation, the $^{60}$Fe-$^{60}$Ni system would record the time of core formation in the building blocks of Mars, not core formation in Mars itself. Such scenarios have been considered for Earth because chaotic growth is thought to take place through the collisions of large bodies (Chambers and Wetherill, 1998, Raymond et al., 2009). For example, the Moon is thought to have formed by the impact with the protoEarth of an embryo the size of Mars or possibly half the size of Earth (Cameron and Ward, 1976; Canup, 2012; Canup and Asphaug, 2001; Ćuk and Stewart, 2012; Hartmann and Davis, 1975; Reufer et al., 2012). The degree to which terrestrial $^{182}$Hf-$^{182}$W isotope systematics is affected by such large impacts is still debated. For Mars, the situation is easier than for Earth as evidence suggests that this planet formed very rapidly from the accretion of small planetesimals (Walsh et al., 2011; Dauphas and Pourmand, 2011; Kobayashi and Dauphas, 2013). Morishima et al. (2013) evaluated how lack of equilibration between impactor-core and target-mantle could influence the accretion timescale of Mars inferred from $^{182}$Hf-$^{182}$W systematics and they concluded that the effect was small and that Mars formed rapidly. Rapid accretion means that there was enough $^{26}$Al for both Mars and the incoming bodies to be extensively molten (Grimm and McSween, 1993; Dauphas and Pourmand, 2011), favoring efficient mixing. Furthermore, the small size of the accreting planetesimals means that they were vaporized upon impact, allowing equilibration with the target mantle (Nimmo and Agnor, 2006). Their small sizes also allowed sinking metal to break down into droplets of size around 20 cm that could have easily equilibrated with the surrounding medium (Rubie et al., 2003; Samuel, 2012; Deguen et al., 2011). Accordingly, we assume complete $^{60}$Fe-$^{60}$Ni equilibration between incoming planetesimals and the mantle of proto-Mars.

Halliday et al. (1996), Harper and Jacobsen (1996), and Jacobsen (2005) developed models to calculate the isotopic evolution of $^{182}$W in planetary bodies for different accretion timescales. The equations given in these papers incorporate the fact that the core does not contain any of the parent nuclide $^{182}$Hf. This assumption is not valid for $^{60}$Fe-$^{60}$Ni systematics because the core of Mars would have been an important repository of $^{60}$Fe and the equations have to be modified accordingly (Appendix A):

$$\varepsilon^{60}\text{Ni}_{\text{Mars mantle}}(t) - \varepsilon^{60}\text{Ni}_{\text{Chondrite}}(t) = \lambda f_m q_{Ni} \left(\frac{^{60}\text{Fe}}{^{56}\text{Fe}}\right)_{\text{Chondrite},0} \int_0^t \left[\frac{M(x)}{M(t)}\right]^{1-\frac{f_m}{f_c}} e^{-\lambda x} dx \quad (6)$$

Using the parameterization for embryo growth given in Eq. 5, this equation takes the form,

$$\varepsilon^{60}\text{Ni}_{\text{Mars mantle}}(t) - \varepsilon^{60}\text{Ni}_{\text{Chondrite}}(t)$$

$$= \lambda f_m q_{Ni} \left(\frac{^{60}\text{Fe}}{^{56}\text{Fe}}\right)_{\text{Chondrite},0} \int_0^t \left[\frac{\tanh(x/\tau)}{\tanh(t/\tau)}\right]^{3(1-\frac{f_m}{f_c})} e^{-\lambda x} dx \quad (7)$$

Where $q_{Ni} = 10^4 (^{56}\text{Fe}/^{60}\text{Ni})_{\text{Chondrite}} = 595525$ (Wasson and Kallemeyn, 1988) and $\varepsilon^{60}\text{Ni}_{\text{Chondrite}}(\text{present}) = -0.038 \pm 0.006$ (this study, Steele et al., 2012; Tang and Dauphas, 2012) assuming that Mars is made of 55 % OC+ 45 % EC; $(^{60}\text{Fe}/^{56}\text{Fe})_{\text{Chondrite},0} = (1.15 \pm 0.26) \times 10^{-8}$ at CAI formation (Tang and Dauphas, 2012); $\varepsilon^{60}\text{Ni}_{\text{Mars mantle}}(\text{present}) = -0.010 \pm 0.022$ (this study; Table 1); $\lambda = 0.26456$ Myr$^{-1}$ is the decay constant of $^{60}$Fe; $f_m = (\text{Fe/Ni})_{\text{Mars mantle}}/(\text{Fe/Ni})_{\text{Chondrite}} - 1 = 19.38$; $f_c = (\text{Fe/Ni})_{\text{Mars core}}/(\text{Fe/Ni})_{\text{Chondrite}} - 1 = -0.430$.

In Fig. 5, the difference $\varepsilon^{60}\text{Ni}_{\text{Mars mantle}}(t) - \varepsilon^{60}\text{Ni}_{\text{Chondrite}}(t)$ is plotted as a function of time for different accretion timescales. If Mars had formed very rapidly at the time of CAI formation ($\tau=0$ Myr), then excess $\varepsilon^{60}$Ni of the bulk martian mantle relative to chondrites should have been +0.14. Conversely, if Mars' core had formed after complete decay of $^{60}$Fe, then there should not have been any excess $^{60}$Ni. The measured difference $\varepsilon^{60}\text{Ni}_{\text{Mars mantle}}(\text{present}) - \varepsilon^{60}\text{Ni}_{\text{Chondrite}}(\text{present}) = +0.028\pm0.023$ translates into an accretion timescale of $1.9^{+1.7}_{-0.8}$ Myr. The error bar on $\tau$ is the 95 % confidence interval calculated by propagating uncertainties on all model parameters using a Monte-Carlo approach. The upper-bound is loosely defined because the $\varepsilon^{60}$Ni value of SNC meteorites is barely resolvable from that of chondrites. However, the lower bound of 1.2 Myr after CAI formation is a firm limit. Indeed, any accretion timescale lower than that would have produced enough radiogenic $^{60}$Ni in the martian mantle to be readily detected by our technique.

The timescale obtained for Mars core formation and accretion of $1.9^{+1.7}_{-0.8}$ Myr after CAI formation agrees with the previous value of $1.8^{+0.9}_{-1.0}$ Myr based on $^{182}$Hf-$^{182}$W systematics of SNC meteorites (Dauphas and Pourmand, 2011). As discussed previously, the timescale obtained using $^{182}$Hf-$^{182}$W systematics is strictly speaking an upper-limit. On the other hand, $^{60}$Fe-$^{60}$Ni systematics provides a solid lower limit on the accretion timescale of Mars. Combining the two estimates, we obtain $1.8^{+0.8}_{-0.6}$ Myr. This means that Mars had reached approximately 44 % of its present size by ~1.8 Myr, a time when $^{26}$Al was still alive and provided sufficient heat to melt planetary bodies including Mars.

The new constraint on the accretion timescale of Mars can help put its formation in the broader context of early solar system evolution. Application of $^{182}$Hf-$^{182}$W systematics to magmatic iron meteorites indicates that most of them were segregated as metallic cores from their parent-bodies within approximately ~1 Myr of the formation of the solar system (Kleine et al., 2005, 2009; Schersten et al., 2006; Markowski et al., 2006; Qin et al., 2008; Burkhardt et al., 2012; Kruijer et al., 2012; Wittig et al., 2013). At that time,

$^{26}$Al was a potent heat source, explaining why all the remnants from this early stage of planetesimal formation are differentiated (*i.e.*, molten) meteorites. Planetesimal formation was a protracted phenomenon that continued at least until ~3 Myr after CAI formation (Kita et al., 2005; Villeneuve et al., 2009; Dauphas and Chaussidon 2011 and references therein). The planetesimals from that period are mostly undifferentiated because $^{26}$Al had decayed to a too low level to induce significant melting. A conclusion of the present study is that Mars would have accreted while planetesimals were still forming. By the time of chondrite formation, Mars would have already reached almost its full size, raising the possibility of forming chondrules by bow shocks generated by embryos (Morris et al., 2012; Boley et al., 2013) or by collisions between planetesimals (Asphaug et al., 2011; Sanders and Scott, 2012; Fedkin and Grossman, 2013) that would have been dynamically excited by the presence of an embryo.

A similar approach to that used for Mars or Vesta cannot be applied to Earth because its accretion timescale was undoubtedly too long for $^{60}$Fe decay to impart any variation in the isotopic composition of $^{60}$Ni in the terrestrial mantle. Furthermore, at high pressure and high temperature Ni tends to become more lithophile (Thibault and Walter, 1995; Li and Agee, 1996; Ito et al., 1998), so the Fe/Ni ratio in Earth's mantle is too close to the chondritic ratio to yield any useful constraints on the accretion timescale of Earth.

5. Conclusion

The $^{60}$Fe-$^{60}$Ni extinct radionuclide system can provide constraints on the timing of core formation in early-formed planetary bodies. Using a new estimate of the initial $^{60}$Fe/$^{56}$Fe ratio of $(1.15 \pm 0.26) \times 10^{-8}$ in the solar system, we present the first chronological application of the $^{60}$Fe-$^{60}$Ni decay system to establish the timescale of accretion and core segregation in Mars. The inferred timescale of $1.9^{+1.7}_{-0.8}$ Myr after condensation of the first solids agrees with a value of $1.8^{+0.9}_{-1.0}$ Myr obtained using the $^{182}$Hf-$^{182}$W decay system. However, this last estimate was strictly speaking an upper-limit, while $^{60}$Fe-$^{60}$Ni provides a very robust lower limit. The two approaches are thus complementary.

The short accretion timescale obtained for Mars indicates that it grew while planetesimals were still being accreted. Most simulations of embryo growth start by considering a disk populated by already formed planetesimals, an assumption that should be relaxed in light of our results. The presence of an embryo when planetesimals were still being formed may have influenced the formation of chondrules through bow shocks or by inducing collisions between dynamically excited planetesimals.

Constraints on the accretion timescale of planets are hard to come by and the successful application of $^{60}$Fe-$^{60}$Ni to Mars accretion and differentiation is a major advance in our understanding of early solar system chronology.

**Acknowledgements.**

This work was supported by grants from the NASA Cosmochemistry (NNX12AH60G) and NSF (EAR1144429) programs. The samples were generously provided by the Robert A. Pritzker Center for Meteoritics at the Field Museum (Philipp R. Heck) and Smithsonian National Museum of Natural History (Timothy J. McCoy). Discussions with Hiroshi Kobayashi, Alessandro Morbidelli, Mark Fornace, and Robert

**Figure caption:**

**Fig. 1.** Nickel isotope compositions of iron and chondritic meteorites analysed by Steele et al. (2012) (open symbols), Dauphas et al. (2008) and this study (filled symbols).

**Fig. 2.** Isotopic anomalies in chondrites and SNC meteorites: (A) $\varepsilon^{62}$Ni *vs.* $\varepsilon^{54}$Cr; (B) $\varepsilon^{62}$Ni *vs.* $\Delta^{17}$O, (C) $\varepsilon^{62}$Ni *vs.* $\varepsilon^{50}$Ti, and (D) $\varepsilon^{62}$Ni *vs.* $\varepsilon^{92}$Mo (Cr: Qin et al., 2010; Trinquier et al., 2007; Ti: Trinquier et al. 2009; Zhang et al., 2012; Ni: Steele et al., 2012; Tang and Dauphas, 2012; O: Clayton et al., 1983, 1984, 1991, 1999; Weisberg et al., 2001; Mo: Dauphas et al., 2002a, b; Burkhardt et al., 2011). Proposed model compositions of bulk Mars are also shown (85 % H+11 % CV+4 % CI; Lodders and Fegley, 1997; 55 % H+45 % EH, Sanloup et al., 1999). Only the model of Sanloup et al. (1999) can reproduce the isotopic anomalies measured in SNC meteorites. In all the panels, terrestrial composition is at (0,0) coordinate.

**Fig. 3.** $\varepsilon^{60}$Ni values in terrestrial standards (black cycles), chondrites (diamonds), and Martian meteorites (open squares) (Table 1; Steele et al., 2012; Dauphas et al., 2008; Tang and Dauphas, 2012). Ni isotopic ratios are reported in the $\varepsilon$-notation; $\varepsilon^{60}$Ni=[$(^{60}$Ni/$^{58}$Ni$)_{sample}$/$(^{60}$Ni/$^{58}$Ni$)_{standard}$-1]×10$^4$, where $^{60}$Ni/$^{58}$Ni ratios have been corrected for natural and laboratory-introduced mass fractionation by internal normalization to a constant $^{61}$Ni/$^{58}$Ni ratio. The error bars represent 95 % confidence intervals. The gray bars represent the weighted averages of chondrites. According to the mixing model described in the text (Sanloup et al., 1999), SNC values for $\Delta^{17}$O, $\varepsilon^{50}$Ti, $\varepsilon^{54}$Cr, $\varepsilon^{62}$Ni, and $\varepsilon^{92}$Mo can be reproduced with 55% ordinary chondrites (OC), and 45% enstatite chondrites (EC), corresponding to a bulk Mars $\varepsilon^{60}$Ni value of -0.038±0.006. All

SNC meteorites analyzed in this study have terrestrial $\varepsilon^{60}$Ni values, averaging -0.010±0.022 (red bar and red square). This is taken to be the isotopic composition of the Martian mantle (see text for details).

**Fig. 4.** $^{60}$Fe-$^{60}$Ni diagram of Martian meteorites (see Fig. 3 caption for notations). In $\varepsilon^{60}$Ni *vs.* $^{56}$Fe/$^{58}$Ni isochron diagram, the intercept gives the initial Ni isotopic composition $\varepsilon^{60}$Ni$_0$, while the slope is proportional to the initial $^{60}$Fe/$^{56}$Fe ratio; slope=25,961×($^{60}$Fe/$^{56}$Fe)$_0$. No correlation was found between $\varepsilon^{60}$Ni *vs.* $^{56}$Fe/$^{58}$Ni in Martian meteorites (open squares). The average $\varepsilon^{60}$Ni (-0.010±0.022) value in SNC meteorites is taken to be representative of the Martian mantle (red square). The model isochron defined by the bulk martian mantle (SNC meteorites, red square) and bulk Mars (chondrites, green dot) gives an initial $^{60}$Fe/$^{56}$Fe ratio of $2.5 \times 10^{-5}$.

**Fig. 5.** $\varepsilon^{60}$Ni isotope evolution of the mantle of Mars ($^{58}$Fe/$^{56}$Ni = 472; Warren et al., 1999) for different accretion timescales ($\tau$) of Mars; $M_{Mars}=M_{Final} \times \tanh^3(t/\tau)$. The x-axis is the time (t) after solar system formation defined as CAI condensation. $\varepsilon^{60}$Ni isotopic composition in Martian mantle remains constant after 10 Myr within our uncertainty due to complete decay of $^{60}$Fe. The Ni isotopic composition of SNC meteorites (red dot) constrains the accretion timescale of Mars to be $1.9^{+1.7}_{-0.8}$ Myr.

**Appendix A: Modeling the $^{60}$Ni isotopic evolution of the mantle of a growing planet**

The following notations are used:

- $\varepsilon^{60}\text{Ni}_i = [(^{60}\text{Ni}/^{58}\text{Ni})_i/(^{60}\text{Ni}/^{58}\text{Ni})_{std} - 1] \times 10{,}000$,

- $M_m$, $M_c$, and $M = M_m + M_c$ are the masses of the mantle, core, and whole planet, respectively,

- $[E]_m$, $[E]_c$, and $[E]_{m \to c}$ are the concentrations of element/isotope E in the mantle, core, and mass-flux from mantle to core, respectively,

- $\gamma = M_c/(M_m + M_c)$,

- $D^E = [E]_{metal}/[E]_{silicate}$,

- $f_m = (\text{Fe}/\text{Ni})_m/(\text{Fe}/\text{Ni})_{CHUR} - 1$ and $f_c = (\text{Fe}/\text{Ni})_c/(\text{Fe}/\text{Ni})_{CHUR} - 1$,

- $R = {}^{60}\text{Ni}/{}^{58}\text{Ni}$.

We start by writing the following equation governing the concentration of Fe in the mantle,

$$d(M_m[\text{Fe}]_m) = [\text{Fe}]_{CHUR} d(M_m + M_c) - [\text{Fe}]_{m \to c} dM_c. \quad (A1)$$

The metal that is removed is in equilibrium with the silicate at each time step and we have,

$$(1-\gamma)d(M[\text{Fe}]_m) = [\text{Fe}]_{CHUR} dM - \gamma D^{Fe}[\text{Fe}]_m dM. \quad (A2)$$

After some rearrangement, it follows,

$$d[\text{Fe}]_m = \frac{1}{1-\gamma}\{[\text{Fe}]_{CHUR} - (1 - \gamma + \gamma D^{Fe})[\text{Fe}]_m\} d\ln M. \quad (A3)$$

With the initial condition that at the beginning of accretion, bulk metal must be in equilibrium with bulk silicate, we find that the concentrations in the mantle (and by mass-balance the core) must remain constant throughout accretion if $\gamma$ and $D^{Fe}$ remain constant,

$$[Fe]_m = [Fe]_{CHUR}/(1 - \gamma + \gamma D^{Fe}), \quad (A4)$$

$$[Fe]_c = D^{Fe}[Fe]_{CHUR}/(1 - \gamma + \gamma D^{Fe}). \quad (A5)$$

Although the expressions for these concentrations correspond to equilibrium values, it does not mean that the bulk core is in equilibrium with the bulk mantle. Similar equations apply to Ni and the $f_m$ and $f_c$ values also remain constant. We can write the following mass balance equation for Ni and Fe between the mantle, core, and bulk planet assumed to be chondritic,

$$[Ni]_c M_c + [Ni]_m M_m = [Ni]_{CHUR}(M_m + M_c), \quad (A6)$$

$$(Fe/Ni)_c[Ni]_c M_c + (Fe/Ni)_m[Ni]_m M_m = (Fe/Ni)_{CHUR}[Ni]_{CHUR}(M_m + M_c). \quad (A7)$$

If we form the difference (A7)-(A6) and divide by $(Fe/Ni)_{CHUR}$, we obtain the relationship,

$$f_c[Ni]_c M_c + f_m[Ni]_m M_m = 0. \quad (A8)$$

We therefore have,

$$\frac{[Ni]_c M_c}{[Ni]_m M_m} = -\frac{f_m}{f_c}, \quad (A9)$$

and similarly,

$$\frac{[Ni]_c M_c + [Ni]_m M_m}{[Ni]_m M_m} = \frac{[Ni]_{CHUR} M}{[Ni]_m M_m} = 1 - \frac{f_m}{f_c}. \quad (A10)$$

We can now write a differential equation for $^{60}$Ni in the mantle,

$$\underset{\text{change in the mantle}}{d(M_m[^{58}Ni]_m R_m)} = \underset{\text{extraterrestrial delivery}}{[^{58}Ni]_{CHUR} R_{CHUR} d(M_m + M_c)} - \underset{\text{removal to the core}}{[^{58}Ni]_c R_m dM_c} + \underset{\text{radioactive decay}}{\lambda M_m [^{60}Fe]_m dt}. \quad (A11)$$

Here it is assumed that the impactor is well-mixed and homogenized in the mantle, so that each increment of metal that is removed to the core has the isotopic composition of the mantle. We have just seen that the concentration in the mantle remains constant, so $[Ni]_m$ in the left term can be taken out of the differential,

$$[^{58}Ni]_m d(M_m R_m) = [^{58}Ni]_{CHUR} R_{CHUR} dM - [^{58}Ni]_c R_m dM_c + \lambda M_m [^{60}Fe]_m dt, \quad (A12)$$

$$[^{58}Ni]_m R_m dM_m + [^{58}Ni]_m M_m dR_m = [^{58}Ni]_{CHUR} R_{CHUR} dM - [^{58}Ni]_c R_m dM_c + \lambda M_m [^{60}Fe]_m dt. \quad (A13)$$

All the terms can be divided by $[^{58}Ni]_m M_m$ to yield,

$$R_m d\ln M_m + dR_m = \frac{[^{58}Ni]_{CHUR} M}{[^{58}Ni]_m M_m} R_{CHUR} d\ln(M) - \frac{[^{58}Ni]_c M_c}{[^{58}Ni]_m M_m} R_m d\ln(M_c) + \lambda \left(\frac{^{60}Fe}{^{58}Ni}\right)_m dt. \quad (A14)$$

Because we assume $\gamma$=constant, we have $d\ln(M_c)=d\ln(M_m)=d\ln(M)$. Injecting Eqs. A9 and A10 into Eq. A14, it follows,

$$dR_m = -R_m d\ln(M) + \left(1 - \frac{f_m}{f_c}\right) R_{CHUR} d\ln(M) + \frac{f_m}{f_c} R_m d\ln(M) + \lambda \left(\frac{^{60}Fe}{^{58}Ni}\right)_m dt, \quad (A15)$$

$$dR_m = \left(\frac{f_m}{f_c} - 1\right)(R_m - R_{CHUR}) d\ln(M) + \lambda(f_m + 1)\left(\frac{^{60}Fe}{^{58}Ni}\right)_{CHUR} dt. \quad (A16)$$

The differential equation governing the change in Ni isotopic composition of the CHUR reservoir is,

$$dR_{CHUR} = \lambda \left(\frac{^{60}Fe}{^{58}Ni}\right)_{CHUR} dt. \quad (A17)$$

Eq. A16 can therefore be rewritten as,

$$d(R_m - R_{CHUR}) = \left(\frac{f_m}{f_c} - 1\right)(R_m - R_{CHUR})d\ln(M) + \lambda f_m \left(\frac{^{60}Fe}{^{58}Ni}\right)_{CHUR} dt. \quad (A18)$$

If we pose $X = R_m - R_{CHUR}$ and express the $(^{60}Fe/^{58}Ni)_{CHUR}$ ratio as a function of the ratio at the time of CAI formation, we have,

$$dX = \left(\frac{f_m}{f_c} - 1\right)X d\ln(M) + \lambda f_m \left(\frac{^{60}Fe}{^{58}Ni}\right)_{CHUR,0} e^{-\lambda t} dt. \quad (A19)$$

This equation can be solved analytically using the initial condition $X(t=0) = 0$,

$$R_m(t) - R_{CHUR}(t) = \lambda f_m \left(\frac{^{60}Fe}{^{58}Ni}\right)_{CHUR,0} \int_0^t \left[\frac{M(x)}{M(t)}\right]^{1-\frac{f_m}{f_c}} e^{-\lambda x} dx. \quad (A20)$$

In $\varepsilon$ notation, this takes the form,

$$\varepsilon^{60}Ni_m(t) - \varepsilon^{60}Ni_{CHUR}(t) = \lambda f_m 10^4 \left(\frac{^{56}Fe}{^{60}Ni}\right)_{CHUR} \left(\frac{^{60}Fe}{^{56}Fe}\right)_{CHUR,0} \int_0^t \left[\frac{M(x)}{M(t)}\right]^{1-\frac{f_m}{f_c}} e^{-\lambda x} dx. \quad (A21)$$

If we introduce $q_{Ni} = 10^4 \left(\frac{^{56}Fe}{^{60}Ni}\right)_{CHUR}$, we have,

$$\varepsilon^{60}Ni_m(t) - \varepsilon^{60}Ni_{CHUR}(t) = \lambda f_m q_{Ni} \left(\frac{^{60}Fe}{^{56}Fe}\right)_{CHUR,0} \int_0^t \left[\frac{M(x)}{M(t)}\right]^{1-\frac{f_m}{f_c}} e^{-\lambda x} dx. \quad (A22)$$

No core sample is usually available to measure $f_c$ but this quantity can be estimated by combining estimates of the mantle and CHUR (bulk planet) compositions (Eq. A10),

$$f_c = \frac{[Ni]_m M_m}{[Ni]_m M_m - [Ni]_{CHUR} M} f_m. \quad (A23)$$

**Appendix B. Supplementary materials**

Supplementary data associated with this article can be found in the online version at …

**Table 1.** Fe/Ni ratios and Ni isotope data of geostandards, chondrites, and martian meteorites.

| Name | Type | Mass (mg) | Fe/Ni (wt.) | $^{56}$Fe/$^{58}$Ni (at.) | Norm. $^{61}$Ni/$^{58}$Ni | | | Norm. $^{62}$Ni/$^{58}$Ni | | |
|---|---|---|---|---|---|---|---|---|---|---|
| | | | | | $\varepsilon^{60}$Ni | $\varepsilon^{62}$Ni | $\varepsilon^{64}$Ni | $\varepsilon^{60}$Ni | $\varepsilon^{61}$Ni | $\varepsilon^{64}$Ni |
| **Terrestrial standards** | | | | | | | | | | |
| BE-N | | 150 | 342 | 477 ±52 | -0.02 ±0.04 | 0.02 ±0.09 | | -0.03 ±0.04 | -0.02 ±0.06 | |
| BHVO-02 | | 164 | 710 | 991 ±80 | -0.05 ±0.06 | -0.11 ±0.17 | | 0.00 ±0.05 | 0.08 ±0.13 | |
| BHVO-02 (2) | | 195 | 727 | 1010 ±30 | 0.01 ±0.12 | -0.04 ±0.09 | | -0.01 ±0.10 | 0.03 ±0.04 | |
| DNC-1 | | 187 | 296 | 413 ±33 | -0.01 ±0.05 | 0.03 ±0.08 | | -0.03 ±0.05 | -0.02 ±0.06 | |
| DNC-1 (2) | | 105 | 258 | 359 ±23 | -0.03 ±0.12 | 0.12 ±0.14 | | -0.06 ±0.09 | -0.11 ±0.20 | |
| DTS-02 | | 20 | 15 | 21 | 0.00 ±0.04 | 0.01 ±0.07 | -0.05 ±0.12 | 0.00 ±0.03 | -0.01 ±0.05 | -0.07 ±0.10 |
| **Weighted Average** | | | | | -0.02 ±0.03 | 0.01 ±0.04 | | -0.03 ±0.03 | 0.00 ±0.02 | |
| **Mean Square Weighted Deviation (MSWD)** | | | | | 0.28 | 1.2 | | 0.119 | 1.19 | |
| **Carbonaceous chondrites** | | | | | | | | | | |
| Orgueil | CI | 12.8 | 17 | 25 | -0.03 ±0.04 | 0.14 ±0.07 | 0.51 ±0.12 | -0.11 ±0.03 | -0.11 ±0.06 | 0.06 ±0.30 |
| Mighei | CM2 | 10.9 | 19 | 26 | -0.10 ±0.04 | 0.14 ±0.10 | 0.55 ±0.22 | -0.17 ±0.06 | -0.11 ±0.07 | 0.34 ±0.14 |
| Murchison | CM2 | 10.1 | 19 | 26 | -0.11 ±0.05 | 0.08 ±0.09 | 0.21 ±0.17 | -0.15 ±0.03 | -0.06 ±0.07 | 0.10 ±0.11 |
| Allende | CV3 | 11.5 | 19 | 26 | -0.09 ±0.06 | 0.14 ±0.09 | 0.39 ±0.17 | -0.17 ±0.06 | -0.11 ±0.07 | 0.18 ±0.10 |
| Vigarano | CV3 | 14.1 | 19 | 26 | -0.10 ±0.03 | 0.14 ±0.08 | 0.35 ±0.13 | -0.17 ±0.04 | -0.11 ±0.06 | 0.14 ±0.10 |
| **Weighted Average** | | | | | -0.09 ±0.03 | 0.13 ±0.04 | 0.39 ±0.14 | -0.14 ±0.04 | -0.10 ±0.03 | 0.17 ±0.11 |
| **MSWD** | | | | | 2.4 | 0.41 | 2.2 | 2.1 | 0.42 | 2.1 |
| **Ordinary chondrites** | | | | | | | | | | |
| Bath | H4 | 7.9 | 15 | 21 | -0.06 ±0.04 | -0.09 ±0.09 | -0.18 ±0.25 | -0.01 ±0.04 | 0.07 ±0.07 | -0.04 ±0.17 |
| Bielokrynitschie | H4 | 14.1 | 15 | 21 | -0.06 ±0.04 | -0.13 ±0.10 | -0.25 ±0.20 | 0.01 ±0.04 | 0.10 ±0.07 | -0.06 ±0.13 |
| Kesen | H4 | 10.7 | 15 | 21 | -0.05 ±0.06 | -0.04 ±0.11 | -0.08 ±0.15 | -0.02 ±0.03 | 0.03 ±0.09 | -0.02 ±0.10 |
| Ochansk | H4 | 20.3 | 15 | 21 | -0.04 ±0.04 | -0.08 ±0.07 | -0.18 ±0.14 | 0.00 ±0.04 | 0.06 ±0.05 | -0.07 ±0.11 |
| Ste. Marguerite | H4 | 11.5 | 15 | 21 | -0.01 ±0.03 | -0.07 ±0.13 | -0.27 ±0.26 | 0.03 ±0.06 | 0.06 ±0.10 | -0.16 ±0.18 |
| Kernouvé | H6 | 31.9 | 15 | 21 | -0.06 ±0.04 | -0.08 ±0.07 | -0.17 ±0.11 | -0.02 ±0.03 | 0.07 ±0.05 | -0.05 ±0.10 |
| **Weighted Average** | | | | | -0.04 ±0.02 | -0.08 ±0.03 | -0.17 ±0.06 | -0.01 ±0.02 | 0.07 ±0.03 | -0.06 ±0.05 |
| **MSWD** | | | | | 1.2 | 0.31 | 0.53 | 0.73 | 0.34 | 0.39 |
| **Martian meteorites** | | | | | | | | | | |
| Chassigny | Chassignites | 111.8 | 430 | 600 ±53 | -0.01 ±0.06 | 0.04 ±0.10 | | -0.03 ±0.03 | -0.03 ±0.07 | |
| Lafayette | Nakhlites | 101.9 | 1696 | 2368 ±210 | 0.00 ±0.05 | 0.05 ±0.12 | | -0.03 ±0.07 | -0.04 ±0.09 | |
| Nakhla | Nakhlites | 97.3 | 1963 | 2741 ±230 | -0.01 ±0.04 | 0.03 ±0.08 | | -0.02 ±0.04 | -0.02 ±0.06 | |
| Shergotty | Shergottites | 160.2 | 2112 | 2949 ±301 | -0.01 ±0.05 | 0.07 ±0.13 | | -0.04 ±0.04 | -0.04 ±0.10 | |
| Zagami | Shergottites | 163.2 | 1577 | 2202 ±226 | -0.02 ±0.05 | 0.01 ±0.10 | | -0.03 ±0.04 | -0.01 ±0.07 | |
| **Weighted Average** | | | | | -0.01 ±0.02 | 0.04 ±0.05 | | -0.03 ±0.02 | 0.03 ±0.03 | |
| **MSWD** | | | | | 0.034 | 0.16 | | 0.13 | 0.108 | |

Note: $\varepsilon^{i}$Ni = ($[^{i}$Ni/$^{58}$Ni$]_{sample}$/$[^{i}$Ni/$^{58}$Ni$]_{SRM986}$-1)×10$^{4}$. The uncertainties are 95% confidence intervals.

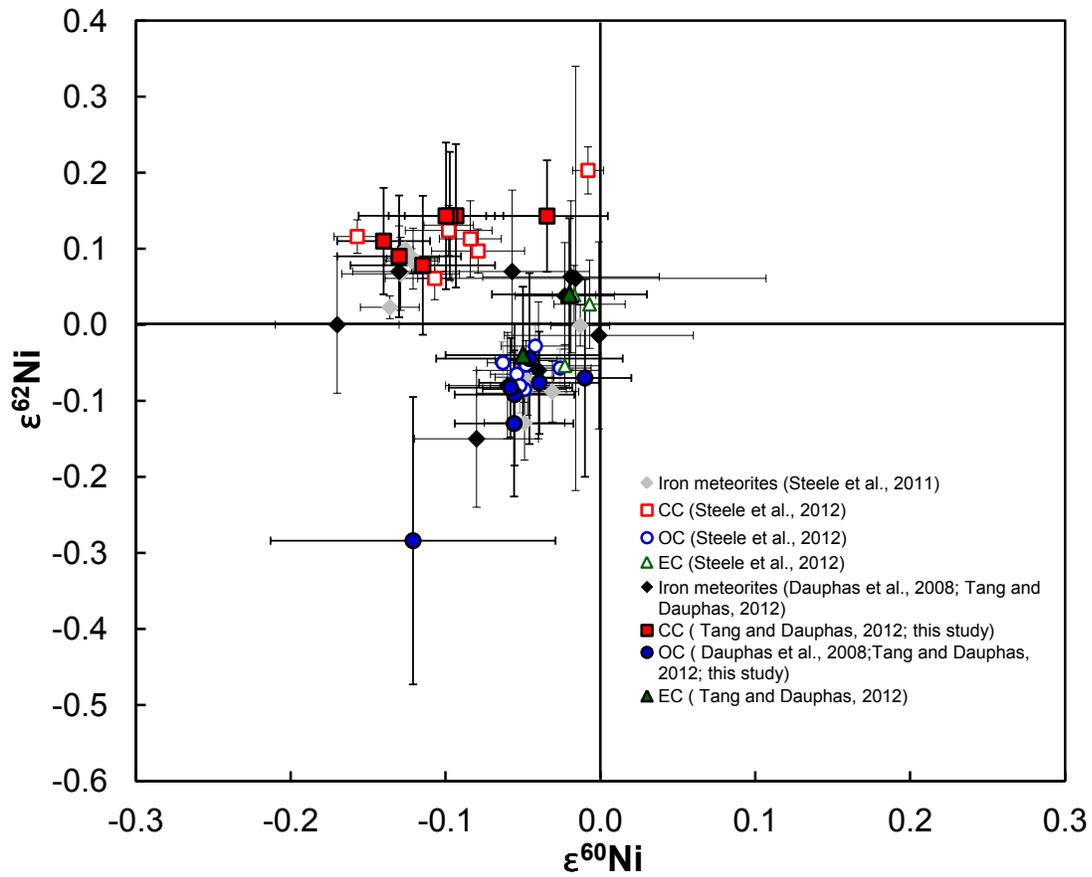
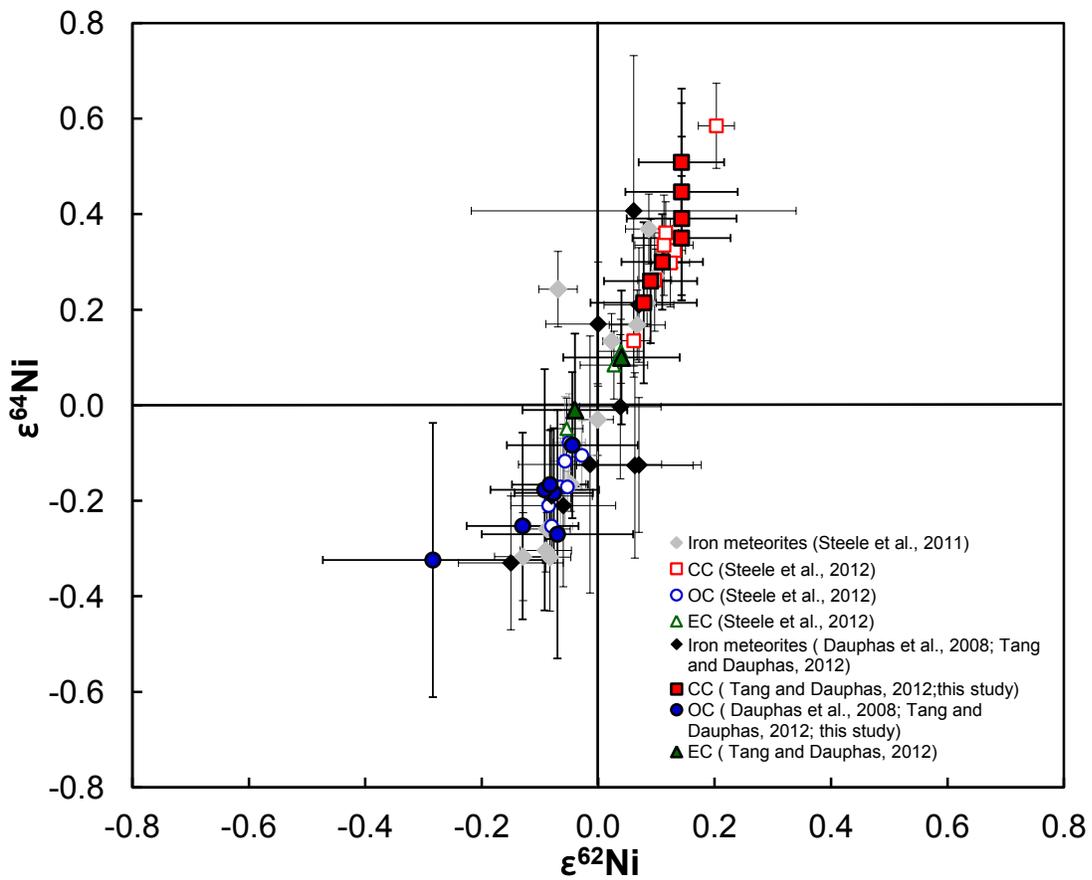

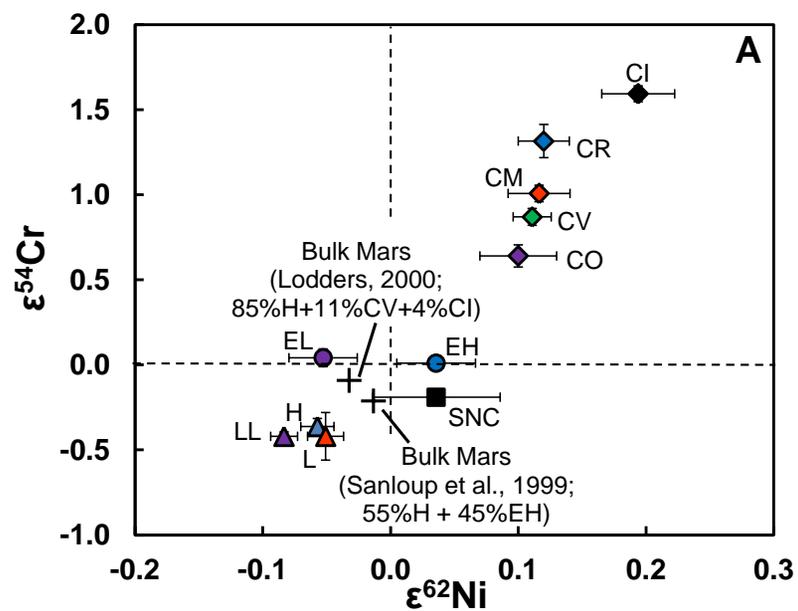
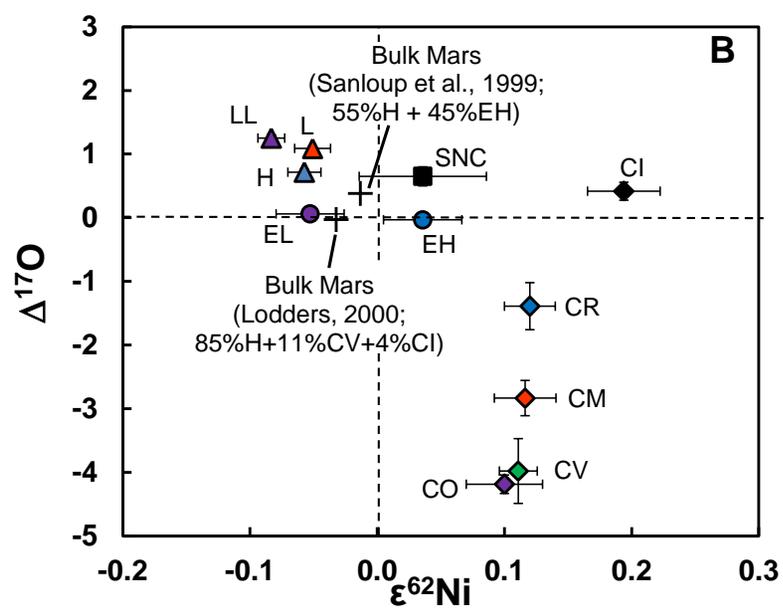
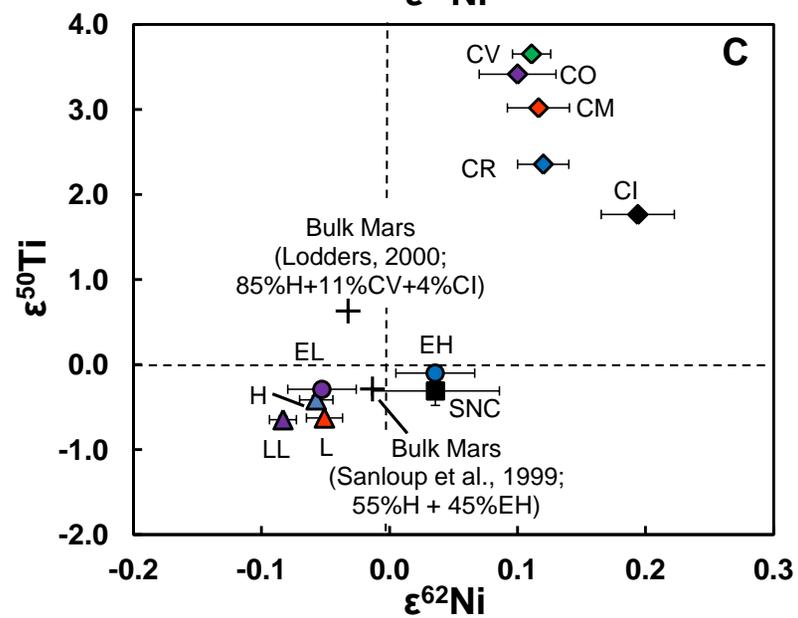
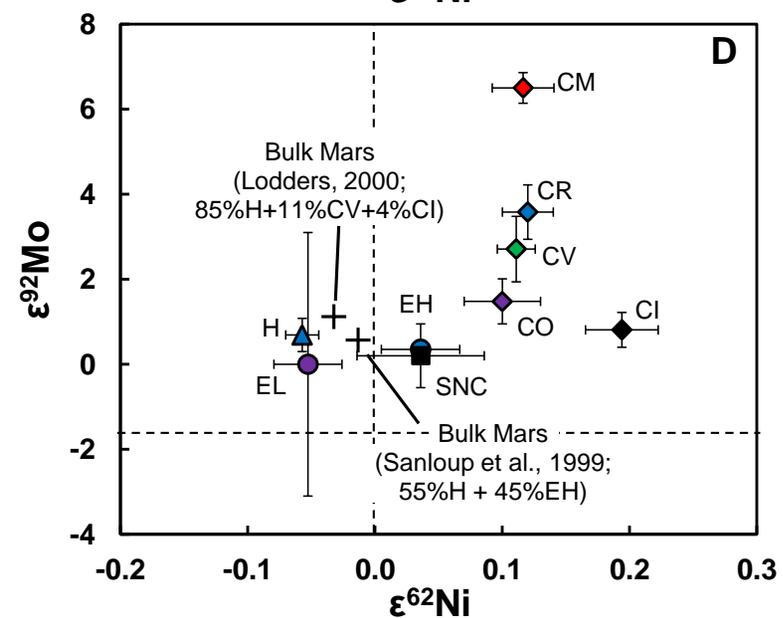

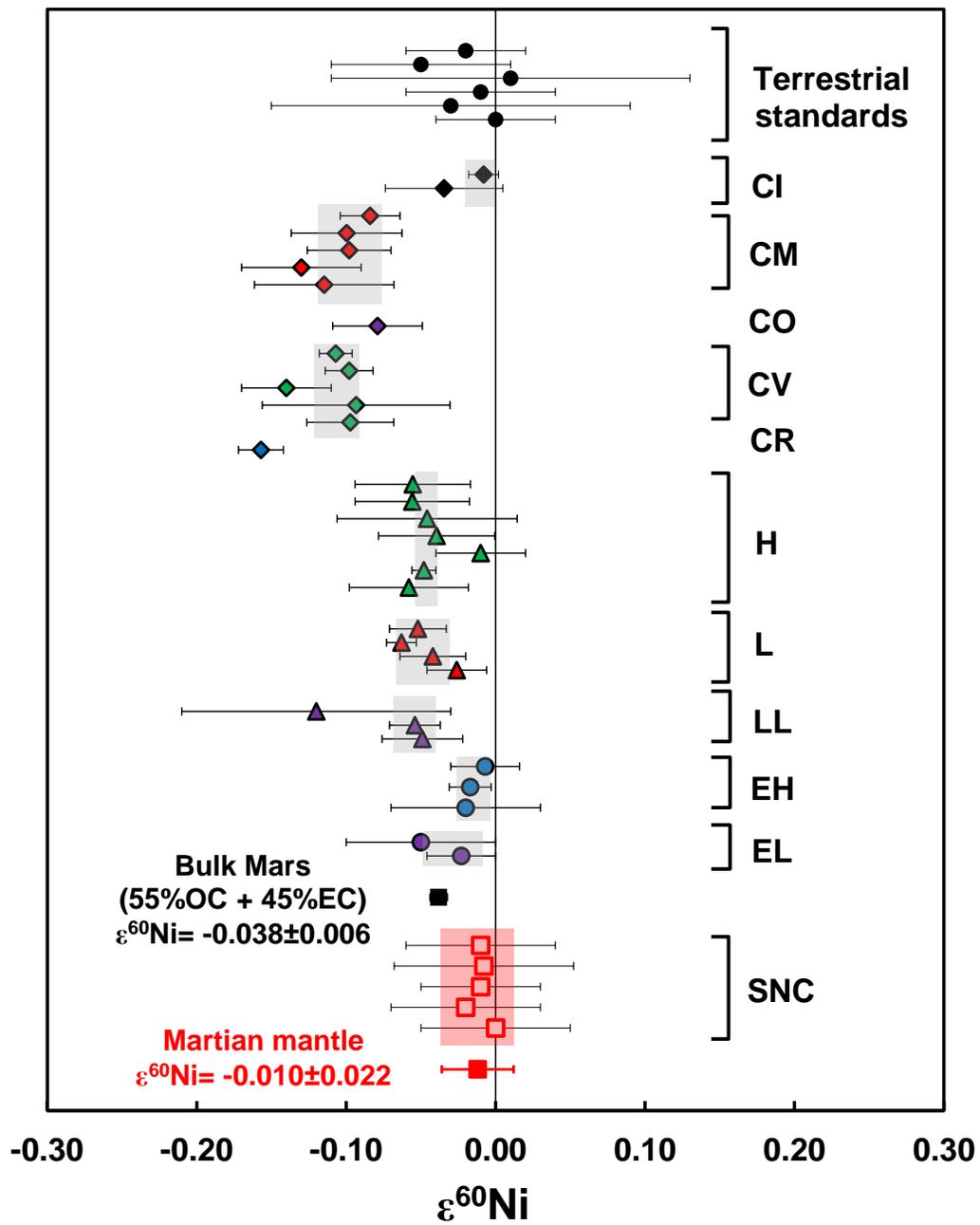

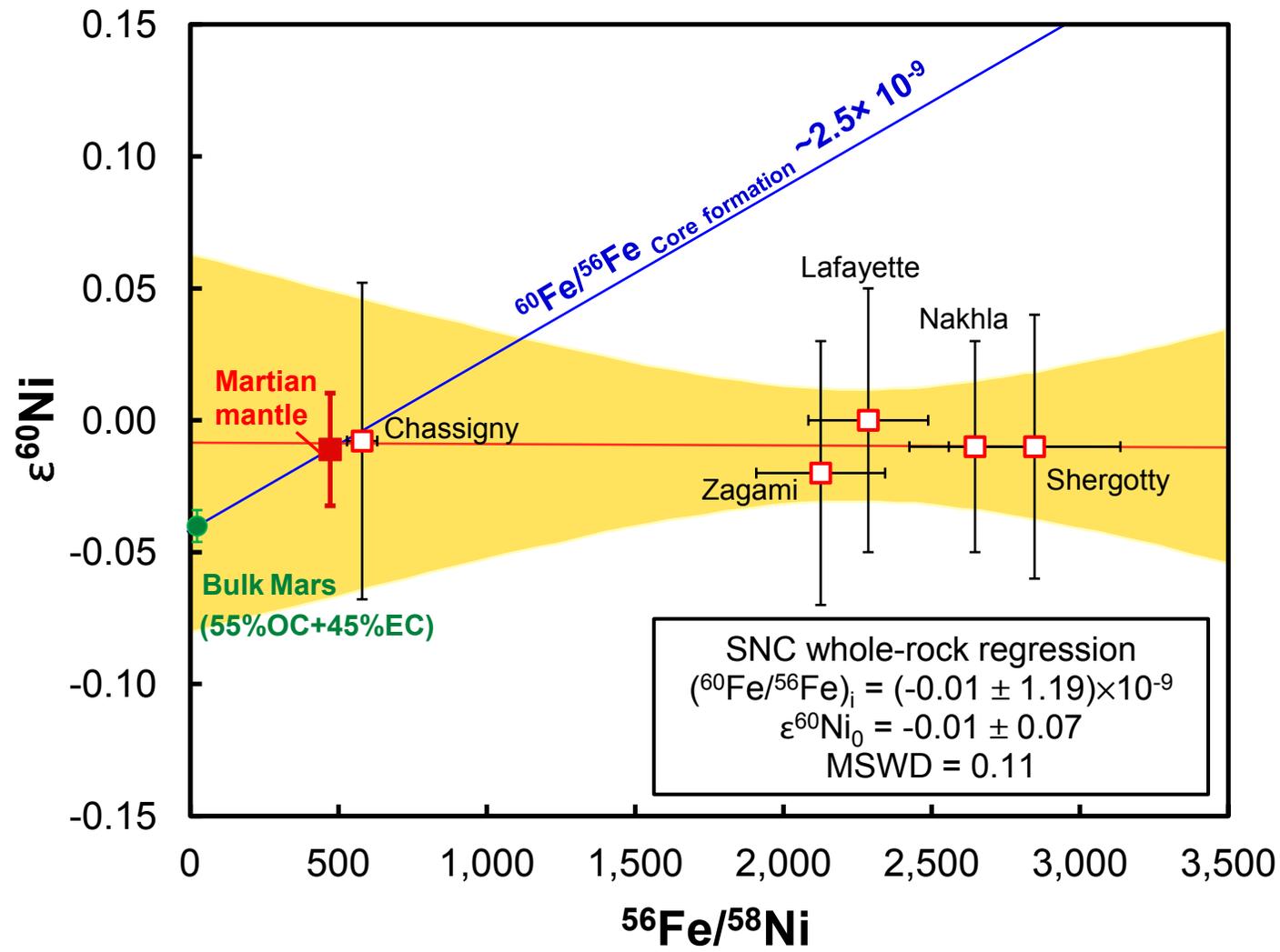

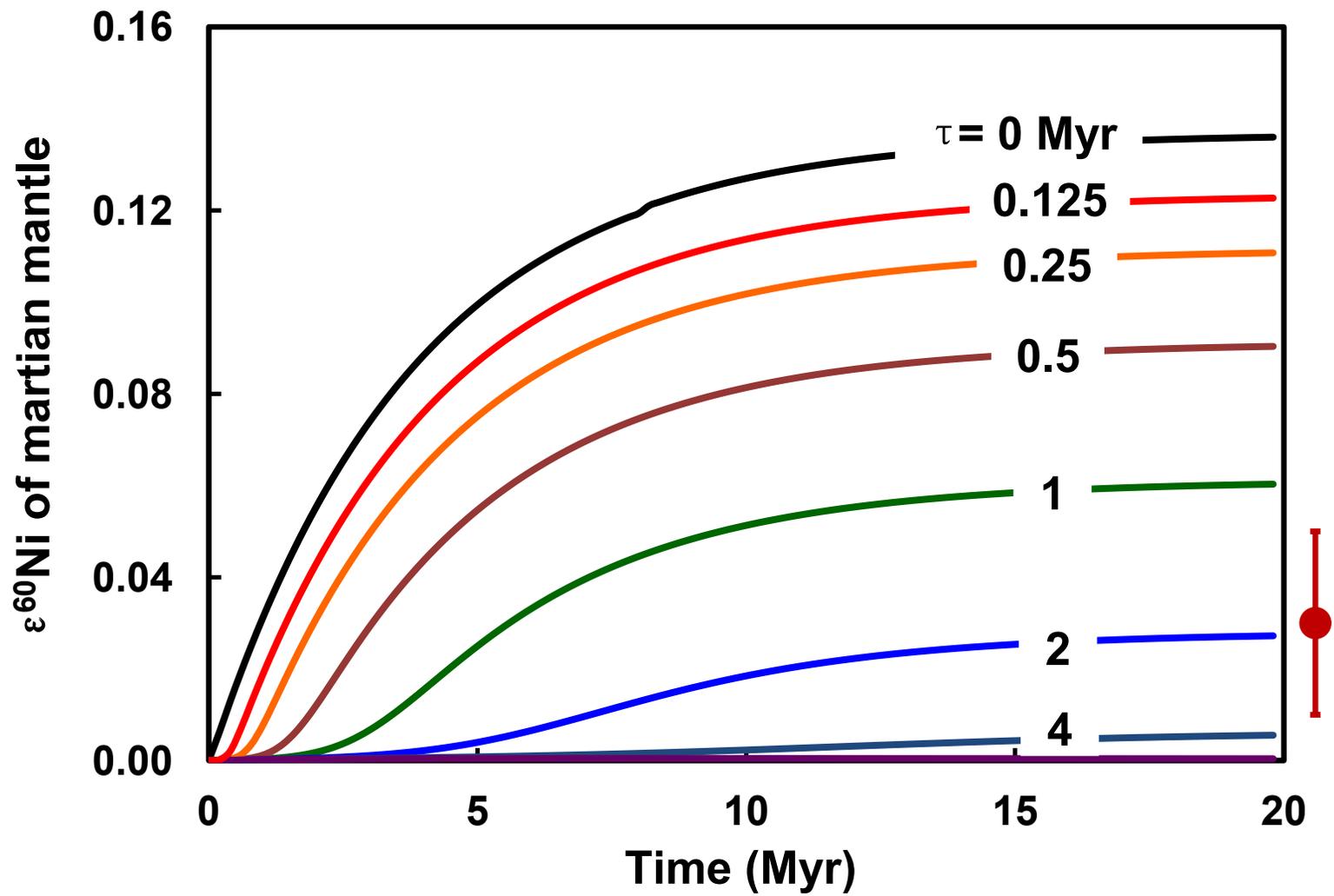

**Supporting Online Material to accompany "$^{60}$Fe-$^{60}$Ni Chronology of Core Formation in Mars" by Haolan Tang and Nicolas Dauphas published in Earth and Planetary Science Letters**

**Table S1.** Nickel isotopic compositions in chondrites and Martian meteorites from previous work and this study.

| Class | Sample | ε$^{60}$Ni | 95% Conf. | ε$^{62}$Ni | 95% Conf. | Reference |
|---|---|---|---|---|---|---|
| **Carbonaceous chondrites** | | | | | | |
| CI | Orgueil | -0.01 | 0.01 | 0.20 | 0.03 | Steele et al., 2012 |
| CI | Orgueil | -0.03 | 0.04 | 0.14 | 0.07 | This study |
| CM2 | Cold-Bokkeveld | -0.08 | 0.02 | 0.11 | 0.05 | Steele et al., 2012 |
| CM2 | Mighei | -0.10 | 0.04 | 0.14 | 0.10 | This study |
| CM2 | Murchison | -0.10 | 0.03 | 0.12 | 0.03 | Steele et al., 2012 |
| CM2 | Murchison | -0.13 | 0.04 | 0.09 | 0.08 | Tang et al., 2012 |
| CM2 | Murchison | -0.11 | 0.05 | 0.08 | 0.09 | This study |
| CO3 | Felix | -0.08 | 0.03 | 0.10 | 0.03 | Steele et al., 2012 |
| CV3 | Leoville | -0.11 | 0.01 | 0.06 | 0.03 | Steele et al., 2012 |
| CV3 | Allende | -0.10 | 0.02 | 0.13 | 0.02 | Steele et al., 2012 |
| CV3 | Allende | -0.14 | 0.03 | 0.11 | 0.07 | Tang et al., 2012 |
| CV3 | Allende | -0.09 | 0.06 | 0.14 | 0.09 | This study |
| CV3 | Vigarano | -0.10 | 0.03 | 0.14 | 0.08 | This study |
| CR2 | NWA-801 | -0.16 | 0.02 | 0.12 | 0.02 | Steele et al., 2012 |
| **Weighted average** | | **-0.10** | **0.02** | **0.12** | **0.02** | |
| **MSWD** | | **29*** | | **4.1*** | | |
| **Ordinary chondrites** | | | | | | |
| H4 | Bath | -0.06 | 0.04 | -0.09 | 0.09 | This study |
| H4 | Bielokrynitschie | -0.06 | 0.04 | -0.13 | 0.10 | This study |
| H4 | Kesen | -0.05 | 0.06 | -0.04 | 0.11 | This study |
| H4 | Ochansk | -0.04 | 0.04 | -0.08 | 0.07 | This study |
| H4 | Ste. Marguerite | -0.01 | 0.03 | -0.07 | 0.13 | This study |
| H6 | Butsura | -0.05 | 0.01 | -0.05 | 0.01 | Steele et al., 2012 |
| H6 | Kernouvé | -0.06 | 0.04 | -0.08 | 0.07 | This study |
| H/L3.6 | Tieschitz | -0.05 | 0.02 | -0.08 | 0.04 | Steele et al., 2012 |
| L3.7 | Ceniceros | -0.06 | 0.01 | -0.05 | 0.03 | Steele et al., 2012 |
| L4 | Barratta | -0.04 | 0.02 | -0.03 | 0.03 | Steele et al., 2012 |
| L6 | Tenham | -0.03 | 0.02 | -0.06 | 0.03 | Steele et al., 2012 |
| LL3.1 | Bishunpur | -0.12 | 0.09 | -0.06 | 0.10 | Dauphas et al., 2008 |
| LL3.4 | Chainpur | -0.05 | 0.02 | -0.07 | 0.04 | Steele et al., 2012 |
| LL6 | Dhurmsala | -0.05 | 0.03 | -0.09 | 0.01 | Steele et al., 2012 |
| **Weighted average** | | **-0.05** | **0.01** | **-0.06** | **0.01** | |
| **MSWD** | | **1.8*** | | **2.2*** | | |
| **Enstatite chondrites** | | | | | | |
| EH4 | Abee | -0.01 | 0.02 | 0.03 | 0.06 | Steele et al., 2012 |
| EH5 | St.Mark's | -0.02 | 0.01 | 0.04 | 0.04 | Steele et al., 2012 |
| EH5 | St.Mark's | -0.02 | 0.05 | 0.04 | 0.10 | Tang et al., 2012 |
| EL5 | Khairpur | -0.05 | 0.05 | -0.04 | 0.09 | Tang et al., 2012 |
| EL6 | Khairpur | -0.02 | 0.02 | -0.05 | 0.03 | Steele et al., 2012 |
| **Weighted average** | | **-0.02** | **0.01** | **0.00** | **0.04** | |
| **MSWD** | | **0.69** | | **4.8*** | | |
| **Martian chondrites** | | | | | | |
| | Shergotty | -0.01 | 0.05 | 0.07 | 0.13 | This study |
| | Chassigny | -0.01 | 0.06 | 0.04 | 0.10 | This study |
| | Nakhla | -0.01 | 0.04 | 0.03 | 0.08 | This study |
| | Zagami | -0.02 | 0.05 | 0.01 | 0.10 | This study |
| | Lafayette | 0.00 | 0.05 | 0.05 | 0.12 | This study |
| **Weighted average** | | **-0.01** | **0.02** | **0.04** | **0.05** | |
| **MSWD** | | **0.034** | | **0.16** | | |

* When the MSWD was significantly higher than 1, the weighted average was calculated using Isoplot by weighting the data by a combination of assigned errors and constant external errors to reduce the value of MSWD (Mean Square Weighted) to 1.

**Table S2.** Compilation of $\Delta^{17}O$ in chondrites and martian meteorites.

| Class | Sample | $\Delta^{17}O$ | Reference | Class | Sample | $\Delta^{17}O$ | Reference | Class | Sample | $\Delta^{17}O$ | Reference |
|---|---|---|---|---|---|---|---|---|---|---|---|
| **Carbonaceous chondrite** | | | | | | | | | | | |
| CI | Alais | 0.39 | Clayton et al., 1999 | CO3 | DaG 025 | -4.24 | Clayton et al., 1999 | CK5 | EET90004 | -4.20 | Clayton et al., 1999 |
| CI | Ivuna | 0.47 | Clayton et al., 1999 | CO3 | Felix | -4.59 | Clayton et al., 1999 | CK5 | Y 82104 | -4.30 | Clayton et al., 1999 |
| CI | Orgueil | 0.39 | Clayton et al., 1999 | CO3 | HH 043 | -4.11 | Clayton et al., 1999 | CK6 | LEW 87009 | -4.32 | Clayton et al., 1999 |
| CM1 | EET 83334 | -2.28 | Clayton et al., 1999 | CO3 | Isna | -4.55 | Clayton et al., 1999 | CR1 | GRO 95577 | -0.45 | Clayton et al., 1999 |
| CM1/2 | ALH 83100 | -2.50 | Clayton et al., 1999 | CO3 | Kainsaz | -4.72 | Clayton et al., 1999 | CR2 | Al Rais | -1.01 | Clayton et al., 1999 |
| CM1/2 | Y 82042 | -2.02 | Clayton et al., 1999 | CO3 | Lance | -4.35 | Clayton et al., 1999 | CR2 | EET 87770 | -1.22 | Clayton et al., 1999 |
| CM2 | A 881334 | -4.16 | Clayton et al., 1999 | CO3 | Ornans | -4.45 | Clayton et al., 1999 | CR2 | El Djouf 001 | -1.47 | Clayton et al., 1999 |
| CM2 | A 881594 | -2.49 | Clayton et al., 1999 | CO3 | Warrenton | -4.07 | Clayton et al., 1999 | CR2 | MAC 87320 | -1.80 | Clayton et al., 1999 |
| CM2 | A 881655 | -2.37 | Clayton et al., 1999 | CO3 | Y 791717 | -4.37 | Clayton et al., 1999 | CR2 | Renazzo | -0.96 | Clayton et al., 1999 |
| CM2 | A 881955 | -2.58 | Clayton et al., 1999 | CV3 | Acfer 082 | -3.73 | Clayton et al., 1999 | CR2 | Y 790112 | -1.56 | Clayton et al., 1999 |
| CM2 | Banten | -2.97 | Clayton et al., 1999 | CV3 | Acfer 086 | -3.40 | Clayton et al., 1999 | CR2 | Y 791498 | -1.64 | Clayton et al., 1999 |
| CM2 | Cimarron | -2.91 | Clayton et al., 1999 | CV3 | Acfer 272 | -2.65 | Clayton et al., 1999 | CR2 | Y 793495 | -1.63 | Clayton et al., 1999 |
| CM2 | Cold Bokkeveld | -2.45 | Clayton et al., 1999 | CV3 | Allende | -3.52 | Clayton et al., 1999 | CR2 | Y 8449 | -2.16 | Clayton et al., 1999 |
| CM2 | EET 87522 | -3.71 | Clayton et al., 1999 | CV3 | Axtell | -3.35 | Clayton et al., 1999 | CH3 | Acfer 182 | -1.55 | Clayton et al., 1999 |
| CM2 | LEW 85311 | -3.23 | Clayton et al., 1999 | CV3 | Ningqiang | -4.55 | Clayton et al., 1999 | CH3 | ALH 85085 | -1.62 | Clayton et al., 1999 |
| CM2 | LEW 87016 | -2.95 | Clayton et al., 1999 | CV3 | Tibooburra | -4.71 | Clayton et al., 1999 | CH3 | PAT 91546 | -1.25 | Clayton et al., 1999 |
| CM2 | LEW 87022 | -2.20 | Clayton et al., 1999 | CV3 | Bali | -2.48 | Clayton et al., 1999 | CH3 | PCA 91328 | -1.50 | Clayton et al., 1999 |
| CM2 | LEW 87148 | -2.15 | Clayton et al., 1999 | CV3 | Bali | -4.90 | Clayton et al., 1999 | CH3 | PCA 91452 | -1.27 | Clayton et al., 1999 |
| CM2 | LEW 88001 | -2.90 | Clayton et al., 1999 | CV3 | Grosnaja | -3.13 | Clayton et al., 1999 | CH3 | PCA 91467 | -1.47 | Clayton et al., 1999 |
| CM2 | LEW 88002 | -4.21 | Clayton et al., 1999 | CV3 | Kaba | -3.50 | Clayton et al., 1999 | CH3 | RKP 92435 | -1.52 | Clayton et al., 1999 |
| CM2 | LEW 90500 | -2.48 | Clayton et al., 1999 | CV3 | Mokoia | -2.74 | Clayton et al., 1999 | CB | Bencubbin | -2.27 | Weisberg et al., 2001 |
| CM2 | MAC 88100 | -2.13 | Clayton et al., 1999 | CV3 | Arch | -4.95 | Clayton et al., 1999 | CB | Weatherford | -2.53 | Weisberg et al., 2001 |
| CM2 | MAC 88101 | -2.24 | Clayton et al., 1999 | CV3 | Efremovka | -5.08 | Clayton et al., 1999 | CB | HH 237 | -2.17 | Weisberg et al., 2001 |
| CM2 | MAC 88176 | -2.03 | Clayton et al., 1999 | CV3 | Leoville | -5.87 | Clayton et al., 1999 | CB | QUE 94411 | -2.25 | Weisberg et al., 2001 |
| CM2 | Mighei | -2.50 | Clayton et al., 1999 | CV3 | Leoville | -3.59 | Clayton et al., 1999 | C1/2-un | Y 82162 | 0.38 | Clayton et al., 1999 |
| CM2 | Murchison | -2.60 | Clayton et al., 1999 | CV3 | QUE 93429 | -5.27 | Clayton et al., 1999 | C2 un | Acfer 094 | -4.52 | Clayton et al., 1999 |
| CM2 | Murray | -3.07 | Clayton et al., 1999 | CV3 | Vigarano | -4.23 | Clayton et al., 1999 | C2 un | Adelaide | -3.62 | Clayton et al., 1999 |
| CM2 | Nogoya | -2.00 | Clayton et al., 1999 | CK3 | DG005 | -4.12 | Clayton et al., 1999 | C2-un | B 7904 | -0.05 | Clayton et al., 1999 |
| CM2 | QUE 93005 | -2.54 | Clayton et al., 1999 | CK3 | Watson 002 | -3.21 | Clayton et al., 1999 | C2-un | Bells | -1.86 | Clayton et al., 1999 |
| CM2 | Y 793321 | -2.94 | Clayton et al., 1999 | CK4 | ALH 82135 | -4.17 | Clayton et al., 1999 | C2-un | Essebi | -1.04 | Clayton et al., 1999 |
| CM2 | Y 793595 | -2.45 | Clayton et al., 1999 | CK4 | Karoonda | -4.48 | Clayton et al., 1999 | C2 un | GRO 95566 | -3.97 | Clayton et al., 1999 |
| CM2 | Y 82054 | -4.84 | Clayton et al., 1999 | CK4 | LEW 86258 | -4.55 | Clayton et al., 1999 | C2 un | MAC 87300 | -3.95 | Clayton et al., 1999 |
| CM2 | Y 82098 | -3.81 | Clayton et al., 1999 | CK4 | LEW 87214 | -4.25 | Clayton et al., 1999 | C2-un | MAC 88107 | -4.39 | Clayton et al., 1999 |
| CM2 | Y 86695 | -4.14 | Clayton et al., 1999 | CK4 | Maralinga | -4.45 | Clayton et al., 1999 | C2-un | Y 86720 | -0.01 | Clayton et al., 1999 |
| CO3 | Acfer 202 | -4.00 | Clayton et al., 1999 | CK4 | PCA 82500 | -3.46 | Clayton et al., 1999 | C2-un | Y 86789 | -0.06 | Clayton et al., 1999 |
| CO3 | Acfer 243 | -4.26 | Clayton et al., 1999 | CK4 | Y 6903 | -4.17 | Clayton et al., 1999 | C3-un | LEW 85332 | -2.42 | Clayton et al., 1999 |
| CO3 | ALH 77307 | -4.47 | Clayton et al., 1999 | CK5 | EET 83311 | -4.32 | Clayton et al., 1999 | C4-un | Coolidge | -4.50 | Clayton et al., 1999 |
| CO3 | DaG 006 | -3.88 | Clayton et al., 1999 | CK5 | EET 87507 | -4.15 | Clayton et al., 1999 | C4-un | HH 073 | -3.76 | Clayton et al., 1999 |
| CO3 | DaG 023 | -4.33 | Clayton et al., 1999 | CK5 | EET 87860 | -4.14 | Clayton et al., 1999 | **Average with 95% confidence** | | **-3.01 ± 0.26** | |

**Table S2. (continued)**

| Class | Sample | $\Delta^{17}O$ | Reference | Class | Sample | $\Delta^{17}O$ | Reference | Class | Sample | $\Delta^{17}O$ | Reference |
|---|---|---|---|---|---|---|---|---|---|---|---|
| **Ordinary chondrite** | | | | | | | | | | | |
| H4 | Bath | 0.71 | Clayton et al., 1991 | L5 | Elenovka | 1.16 | Clayton et al., 1991 | LL6 | Ensisheim | 1.40 | Clayton et al., 1991 |
| H4 | Beaver creek | 0.76 | Clayton et al., 1991 | L5 | Elenovka | 0.97 | Clayton et al., 1991 | LL6 | Jelica | 1.27 | Clayton et al., 1991 |
| H4 | Forest Vale | 0.74 | Clayton et al., 1991 | L5 | Ergheo | 0.98 | Clayton et al., 1991 | LL6 | Mangwandi | 1.14 | Clayton et al., 1991 |
| H4 | Forest Vale | 0.53 | Clayton et al., 1991 | L5 | Farmington | 1.13 | Clayton et al., 1991 | LL6 | St. Mesmin | 1.13 | Clayton et al., 1991 |
| H4 | Kesen | 0.63 | Clayton et al., 1991 | L5 | Homestead | 1.11 | Clayton et al., 1991 | LL6 | St. Séveron | 1.16 | Clayton et al., 1991 |
| H4 | Ochansk | 0.82 | Clayton et al., 1991 | L5 | Knyahinya | 1.05 | Clayton et al., 1991 | LL6 | Vavilovka | 1.21 | Clayton et al., 1991 |
| H4 | Tysnes Island | 0.70 | Clayton et al., 1991 | L5 | Qidong | 1.09 | Clayton et al., 1991 | **Average with 95% confidence** | | **1.01 ± 0.05** | |
| H4 | Weston | 0.88 | Clayton et al., 1991 | L6 | Bachmut | 1.16 | Clayton et al., 1991 | | | | |
| H5 | Allegan | 0.63 | Clayton et al., 1991 | L6 | Barwell | 1.19 | Clayton et al., 1991 | **Enstatite chondrite** | | | |
| H5 | Allegan | 0.59 | Clayton et al., 1991 | L6 | Barwell | 1.14 | Clayton et al., 1991 | EH3 | Qingzhen | -0.03 | Clayton et al., 1984 |
| H5 | Ambapur Nagla | 0.75 | Clayton et al., 1991 | L6 | Bori | 1.06 | Clayton et al., 1991 | EH4 | Adhi Kot | -0.05 | Clayton et al., 1984 |
| H5 | Beardsley | 0.75 | Clayton et al., 1991 | L6 | Cabezo de Mayo | 1.04 | Clayton et al., 1991 | EH4 | Indarch | 0.12 | Clayton et al., 1984 |
| H5 | Forest City | 0.75 | Clayton et al., 1991 | L6 | Mocs | 1.13 | Clayton et al., 1991 | EH4 | Kota Kota | -0.15 | Clayton et al., 1984 |
| H5 | Leighton | 0.87 | Clayton et al., 1991 | L6 | Modoc | 1.08 | Clayton et al., 1991 | EH4 | Parsa | 0.01 | Clayton et al., 1984 |
| H5 | Pantar | 0.56 | Clayton et al., 1991 | L6 | Ramsdof | 0.83 | Clayton et al., 1991 | EH4 | Parsa | 0.05 | Clayton et al., 1984 |
| H5 | Richardton | 0.64 | Clayton et al., 1991 | L6 | Ramsdof | 0.78 | Clayton et al., 1991 | EH4 | South Oman | -0.37 | Clayton et al., 1984 |
| H5 | Richardton | 0.58 | Clayton et al., 1991 | L6 | Tathlith | 1.15 | Clayton et al., 1991 | EH4 | Y691 | -0.05 | Clayton et al., 1984 |
| H5 | Xingyang | 0.81 | Clayton et al., 1991 | L6 | Tillaberi | 0.97 | Clayton et al., 1991 | EH5 | St.Marks | 0.11 | Clayton et al., 1984 |
| H5 | Zhovtnevyi | 0.86 | Clayton et al., 1991 | L6 | Xiujimgin | 1.06 | Clayton et al., 1991 | EH5 | St. Sauveur | 0.04 | Clayton et al., 1984 |
| H6 | Cape Girardeau | 0.72 | Clayton et al., 1991 | LL4 | Albareto | 1.20 | Clayton et al., 1991 | EL6 | Atlanta | 0.03 | Clayton et al., 1984 |
| H6 | Charsonville | 0.65 | Clayton et al., 1991 | LL4 | Bo Xian | 1.15 | Clayton et al., 1991 | EL6 | Blithfield | 0.07 | Clayton et al., 1984 |
| H6 | Guarena | 0.76 | Clayton et al., 1991 | LL4 | Harnler | 1.36 | Clayton et al., 1991 | EL6 | Daniel's Kuil | 0.00 | Clayton et al., 1984 |
| H6 | Kernouve | 0.78 | Clayton et al., 1991 | LL4 | Savtschenskoje | 1.18 | Clayton et al., 1991 | EL6 | Hvittis | 0.07 | Clayton et al., 1984 |
| H6 | Mt. Browne | 0.69 | Clayton et al., 1991 | LL4 | Savtschenskoje | 1.39 | Clayton et al., 1991 | EL6 | Jajh deh Kot Lalu | 0.07 | Clayton et al., 1984 |
| H6 | Queen's Mercy | 0.72 | Clayton et al., 1991 | LL4 | Soko Banja | 1.32 | Clayton et al., 1991 | EL6 | Khairpur | 0.10 | Clayton et al., 1984 |
| L4 | Atarra | 1.14 | Clayton et al., 1991 | LL4 | Witsand Farm | 1.35 | Clayton et al., 1991 | EL6 | Pillistfer | 0.02 | Clayton et al., 1984 |
| L4 | Bjurbole | 1.00 | Clayton et al., 1991 | LL5 | Guidder | 1.19 | Clayton et al., 1991 | EL6 | Ufana | 0.12 | Clayton et al., 1984 |
| L4 | Cynthiana | 1.08 | Clayton et al., 1991 | LL5 | Khanpur | 1.41 | Clayton et al., 1991 | **Average with 95% confidence** | | **0.01 ± 0.06** | |
| L4 | Kendleton | 1.03 | Clayton et al., 1991 | LL5 | Olivernza | 1.14 | Clayton et al., 1991 | | | | |
| L4 | Nan Yang Po | 0.97 | Clayton et al., 1991 | LL5 | Olivernza | 1.11 | Clayton et al., 1991 | **SNC** | | | |
| L4 | Saratov | 1.02 | Clayton et al., 1991 | LL5 | Paragould | 1.39 | Clayton et al., 1991 | shergottite | Shergotty | 0.70 | Clayton et al., 1983 |
| L4 | Tennasilm | 1.15 | Clayton et al., 1991 | LL5 | Siena | 1.32 | Clayton et al., 1991 | shergottite | Zagami | 0.80 | Clayton et al., 1983 |
| L5 | Ausson | 1.12 | Clayton et al., 1991 | LL6 | Appley Bridge | 1.32 | Clayton et al., 1991 | nakhlite | Nakhla | 0.50 | Clayton et al., 1983 |
| L5 | Ausson | 1.16 | Clayton et al., 1991 | LL6 | Appley Bridge | 1.27 | Clayton et al., 1991 | nakhlite | Lafayette | 0.62 | Clayton et al., 1983 |
| L5 | Borkut | 1.06 | Clayton et al., 1991 | LL6 | Dhurmsala | 1.15 | Clayton et al., 1991 | chassignite | Chassigny | 0.62 | Clayton et al., 1983 |
| L5 | Chervettaz | 1.25 | Clayton et al., 1991 | LL6 | Dongtai | 1.02 | Clayton et al., 1991 | **Average with 95% confidence** | | **0.65 ± 0.15** | |

**Table S3.** $^{54}$Cr isotopic compositions in chondrites and Martian meteorites from previous work.

| Class | Sample | ε$^{54}$Cr | 95% Conf. | Reference | Class | Sample | ε$^{54}$Cr | 95% Conf. | Reference |
|---|---|---|---|---|---|---|---|---|---|
| **Carbonaceous chondrites** | | | | | **Enstatite chondrites** | | | | |
| CI | Orgueil | 1.55 | 0.13 | Qin et al., 2010 | EH3 | ALHA 77295 | 0.05 | 0.14 | Qin et al., 2010 |
| CI | Orgueil | 1.69 | 0.09 | Qin et al., 2010 | EH3 | Kota Kota | -0.02 | 0.21 | Trinquier et al., 2007 |
| CI | Orgueil | 1.56 | 0.06 | Trinquier et al., 2007 | EH3 | Kota Kota | 0.04 | 0.07 | Trinquier et al., 2007 |
| CM2 | Murchison | 0.97 | 0.20 | Qin et al., 2010 | EH3 | Qingzhen | -0.02 | 0.08 | Trinquier et al., 2007 |
| CM2 | Murchison | 1.01 | 0.05 | Trinquier et al., 2007 | EH4 | Indarch | 0.05 | 0.14 | Qin et al., 2010 |
| CO3 | Felix | 0.63 | 0.09 | Trinquier et al., 2007 | EH4 | Abee | -0.06 | 0.12 | Trinquieret al., 2007 |
| CO3 | Kainsaz | 0.87 | 0.18 | Qin et al., 2010 | EL3 | MAC 88136 | 0.02 | 0.09 | Qin et al., 2010 |
| CO3 | Lance | 0.57 | 0.11 | Trinquier et al., 2007 | EL6 | Hvittis | -0.01 | 0.17 | Trinquier et al., 2007 |
| CV3 | Allende | 0.98 | 0.14 | Qin et al., 2010 | EL6 | LON 94100 | -0.02 | 0.14 | Qin et al., 2010 |
| CV3 | Allende | 0.92 | 0.13 | Qin et al., 2010 | EL6 | Pillistfer | 0.09 | 0.08 | Trinquier et al., 2007 |
| CV3 | Allende | 0.86 | 0.09 | Trinquier et al., 2007 | **Weighted average** | | **0.02** | **0.03** | |
| CV3 | Leoville | 0.71 | 0.15 | Qin et al., 2010 | **MSWD** | | **0.79** | | |
| CV3 | Vigarano | 0.91 | 0.12 | Qin et al., 2010 | | | | | |
| CV3 | Vigarano | 0.82 | 0.13 | Qin et al., 2010 | **SNC** | | | | |
| CK4 | Karoonda | 0.63 | 0.09 | Trinquier et al., 2007 | | Nakhla | -0.21 | 0.16 | Qin et al., 2010 |
| CR2 | GRA 06100 | 1.32 | 0.11 | Qin et al., 2010 | | Shergotty | -0.18 | 0.17 | Trinquier et al., 2007 |
| CR2 | Renazzo | 1.30 | 0.21 | Trinquier et al., 2007 | | Chassigny | -0.20 | 0.05 | Trinquier et al., 2007 |
| CB | Bencubbin | 1.11 | 0.09 | Trinquier et al., 2007 | | Nakhla | -0.14 | 0.08 | Trinquier et al., 2007 |
| CB | Bencubbin | 1.13 | 0.09 | Trinquier et al., 2007 | | | | | |
| **Weighted average** | | **1.03** | **0.15** | | **Weighted average** | | **-0.19** | **0.04** | |
| **MSWD** | | **50*** | | | **MSWD** | | **0.58** | | |
| | | | | | | | | | |
| **Ordinary chondrites** | | | | | | | | | |
| H4 | Ste Marguerite | -0.39 | 0.07 | Trinquier et al., 2007 | | | | | |
| H4 | LAP 03601 | -0.28 | 0.11 | Qin et al, 2010 | | | | | |
| H6 | Kernouvé | -0.37 | 0.08 | Trinquier et al., 2007 | | | | | |
| L3 | QUE 97008 | -0.42 | 0.14 | Qin et al., 2010 | | | | | |
| LL3.4 | Chainpur | -0.47 | 0.07 | Trinquier et al., 2007 | | | | | |
| LL4 | GRO 95552 | -0.33 | 0.10 | Qin et al., 2010 | | | | | |
| LL6 | Saint Severin | -0.41 | 0.10 | Trinquier et al., 2007 | | | | | |
| LL6 | Saint Severin | -0.42 | 0.03 | Trinquier et al., 2007 | | | | | |
| **Weighted average** | | **-0.41** | **0.04** | | | | | | |
| **MSWD** | | **1.8** | | | | | | | |

* See the footnote of Table S1.

**Table S4.** $^{50}$Ti isotopic compositions in chondrites and Martian meteorites from previous work..

| Class | Sample | $\varepsilon^{50}$Ti | 95% Conf. | Reference | Class | Sample | $\varepsilon^{50}$Ti | 95% Conf. | Reference |
|---|---|---|---|---|---|---|---|---|---|
| **Carbonaceous chondrites** | | | | | **Ordinary chondrites** | | | | |
| CI | Orgueil | 1.92 | 0.27 | Trinquier et al., 2009 | H4 | Kesen | -0.37 | 0.05 | Zhang et al., 2012 |
| CI | Orgueil | 1.86 | 0.11 | Trinquier et al., 2009 | H5 | Juancheng | -0.66 | 0.16 | Trinquier et al., 2009 |
| CI | Orgueil | 1.74 | 0.05 | Zhang et al., 2012 | H6 | St-severin | -0.74 | 0.20 | Trinquier et al., 2009 |
| CM2 | Murchison | 3.32 | 0.16 | Trinquier et al., 2009 | L3.7 | Hedjaz | -0.65 | 0.13 | Trinquier et al., 2009 |
| CM2 | Murchison | 3.06 | 0.33 | Trinquier et al., 2009 | L5 | Ausson | -0.64 | 0.03 | Zhang et al., 2012 |
| CM2 | Murchison | 3.15 | 0.05 | Trinquier et al., 2009 | L6 | Isoulane-n-Amahar | -0.59 | 0.05 | Zhang et al., 2012 |
| CM2 | Murchison | 2.96 | 0.07 | Trinquier et al., 2009 | LL3 | Dar al Gani | -0.72 | 0.16 | Trinquier et al., 2009 |
| CM2 | Murchison | 3.19 | 0.11 | Trinquier et al., 2009 | LL3.2 | Krymka | -0.66 | 0.05 | Zhang et al., 2012 |
| CM2 | Murchison | 2.94 | 0.06 | Trinquier et al., 2009 | LL5 | Paragould | -0.61 | 0.07 | Zhang et al., 2012 |
| CM2 | Murchison | 3.08 | 0.05 | Trinquier et al., 2009 | Un 3 | NWA5717 | -0.58 | 0.05 | Zhang et al., 2012 |
| CM2 | Murray | 2.84 | 0.05 | Zhang et al., 2012 | Un 3 | NWA5717 | -0.63 | 0.05 | Zhang et al., 2012 |
| CO3 | Isna | 3.45 | 0.23 | Trinquier et al., 2009 | **Weight average** | | **-0.61** | **0.06** | |
| CO3 | Lance | 3.46 | 0.10 | Zhang et al., 2012 | **MSWD** | | **10.5*** | | |
| CO3 | Ornans | 3.37 | 0.09 | Zhang et al., 2012 | | | | | |
| CV3 | Allende | 4.98 | 0.29 | Trinquier et al., 2009 | **Enstatite chondrites** | | | | |
| CV3 | Allende | 5.01 | 0.29 | Trinquier et al., 2009 | EH3 | Qingzhen | 0.02 | 0.12 | Trinquier et al., 2009 |
| CV3 | Allende | 3.49 | 0.04 | Zhang et al., 2012 | EH3 | Qingzhen | -0.26 | 0.32 | Trinquier et al., 2009 |
| CV3 | Leoville | 4.09 | 0.08 | Zhang et al., 2012 | EH3 | Sahara 97072 | -0.15 | 0.06 | Zhang et al., 2012 |
| CK4 | Karoonda | 3.26 | 0.10 | Zhang et al., 2012 | EH4 | Indarch | -0.13 | 0.05 | Zhang et al., 2012 |
| CR2 | NWA 801 | 2.35 | 0.04 | Zhang et al., 2012 | EH4 | Adhi Kot | -0.10 | 0.04 | Zhang et al., 2012 |
| CR2 | Sahara 0082 | 2.60 | 0.30 | Trinquier et al., 2009 | EH4 | Abee | -0.06 | 0.04 | Zhang et al., 2012 |
| **Weight average** | | **3.15** | **0.36** | | EH5 | Saint-Sauveur | -0.15 | 0.09 | Zhang et al., 2012 |
| **MSWD** | | **301*** | | | EL6 | Hvittis | -0.29 | 0.07 | Zhang et al., 2012 |
| | | | | | EL6 | Jajh deh Kot Lalu | -0.29 | 0.10 | Zhang et al., 2012 |
| **Martian meteorite** | | | | | **Weight average** | | **-0.15** | **0.07** | |
| SNC | NWA 2737 | **-0.31** | **0.17** | Trinquier et al., 2009 | **MSWD** | | **6.6*** | | |

* See the footnote of Table S1.

**Table S5.** $^{92}$Mo isotopic compositions in chondrites and Martian meteorites from previous work.

| Class | Sample | ε$^{92}$Mo | 95% Conf. | Reference |
|---|---|---|---|---|
| **Carbonaceous chondrites** | | | | |
| CI | Orgueil | 1.12 | 0.59 | Burkhardt et al. 2011 |
| CI | Orgueil | 0.53 | 0.57 | Dauphas et al., 2002 |
| CM2 | Murchison | 6.34 | 1.31 | Burkhardt et al. 2011 |
| CM2 | Murchison | 6.66 | 0.53 | Burkhardt et al. 2011 |
| CM2 | Murchison | 6.36 | 0.55 | Burkhardt et al. 2011 |
| CO3 | NWA2090 | 1.48 | 0.53 | Burkhardt et al. 2011 |
| CV3 | Allende | 3.24 | 0.86 | Burkhardt et al. 2011 |
| CV3 | Allende | 3.46 | 0.51 | Burkhardt et al. 2011 |
| CV3 | Allende | 2.38 | 0.57 | Dauphas et al., 2002 |
| CV3 | Allende | 3.54 | 0.67 | Dauphas et al., 2002 |
| CV3 | Allende | 2.57 | 0.78 | Dauphas et al., 2002 |
| CV3 | Allende | 1.23 | 0.39 | Dauphas et al., 2002 |
| CR2 | NWA801 | 3.58 | 0.64 | Burkhardt et al. 2011 |
| **Weighted average** | | **3.20** | **1.10** | |
| **MSWD** | | **51*** | | |
| **Ordinary chondrites** | | | | |
| H3 | Bremervorde | 0.72 | 0.53 | Burkhardt et al. 2011 |
| H6 | Kernouvé | 0.66 | 0.59 | Burkhardt et al. 2011 |
| **Weighted average** | | **0.69** | **0.39** | |
| **MSWD** | | **0.023** | | |
| **Enstatite chondrites** | | | | |
| EH4 | Abee | 0.65 | 1.11 | Burkhardt et al. 2011 |
| EH4 | Abee | 0.36 | 0.26 | Burkhardt et al. 2011 |
| EH4 | Indarch | -0.44 | 1.25 | Dauphas et al., 2002 |
| EH4 | Indarch | 0.88 | 0.86 | Dauphas et al., 2002 |
| EH4 | Saint Sauveur | 0.15 | 0.62 | Dauphas et al., 2002 |
| EH4 | Saint Sauveur | 0.09 | 1.86 | Dauphas et al., 2002 |
| EL6 | Pilistfer | -1.03 | 0.94 | Dauphas et al., 2002 |
| EL6 | Pilistfer | 1.04 | 1.15 | Dauphas et al., 2002 |
| **Weighted average** | | **0.30** | **0.35** | |
| **MSWD** | | **2.0** | | |
| **Martian meteorites** | | | | |
| | Zagami | 0.52 | 0.75 | Burkhardt et al. 2011 |
| | DaG476 | -0.12 | 0.75 | Burkhardt et al. 2011 |
| **Weighted average** | | **0.20** | **0.75** | |
| **MSWD** | | **1.5** | | |

* See the footnote of Table S1.